\documentclass[prc,unsortedaddress,groupedaddress,preprint,amsmath,amsfonts,amssymb,showpacs,floatfix]{revtex4-1}
\usepackage{graphicx}

\usepackage{amsmath}
\usepackage{bm}
\usepackage{color}

\begin{document}

\title{Surface-integral formalism of deuteron stripping}

\author{A. M. Mukhamedzhanov $^{1}$}
\author{D. Y. Pang$^{2}$}
\author{C. A.  Bertulani$^{3}$}
\author{A. S. Kadyrov$^{4}$}

\affiliation{$^{1}$Cyclotron Institute, Texas A\&M University, College Station, TX 77843, USA }
\affiliation{$^{2}$ School of Physics and Nuclear Energy Engineering, Beihang University, Beijing 100191, China}
\affiliation{$^{3}$Department of Physics, Texas A\&M University-Commerce, Commerce, TX 75429, USA  }
\affiliation{$^{4}$Deptartment of Imaging \& Applied Physics, Curtin University, GPO Box U1987, Perth 6845, Australia}

\date{\today}
                               
\begin{abstract}

The purpose of this paper is to develop  an alternative theory of  deuteron stripping to resonance states based on the surface integral formalism   of Kadyrov et al. [Ann. Phys.  {\bf 324}, 1516 (2009)] and continuum-discretized coupled channels (CDCC).  
 First we demonstrate how the surface integral formalism works in the three-body model and then we consider a more realistic problem  in which a composite structure  of target nuclei is taken via optical potentials.  We explore different choices of channel wave functions and transition operators and show that  a conventional CDCC volume matrix element  can be written in terms of  a surface-integral matrix element, which is peripheral, and  an auxiliary matrix element, which determines the contribution of the nuclear interior over the variable $r_{nA}$.  This auxiliary matrix element appears due to the inconsistency in treating of the $n-A$ potential: this potential should be real in the final state to support  bound states or resonance scattering and  complex in the initial state to describe $n-A$ scattering.  
Our main result is formulation of the  theory of the stripping to resonance states using the prior form of the surface integral formalism and CDCC method.  It is demonstrated that  the conventional CDCC volume matrix element coincides with the surface matrix element, which converges for the stripping to the resonance state. Also the surface representation (over the variable ${\rm {\bf r}}_{nA}$) of the stripping matrix element enhances the peripheral part of the amplitude although the internal contribution doesn't disappear  and increases with increase of the deuteron energy. 
We present calculations corroborating our findings for both stripping to the bound state and the resonance. 

\end{abstract}
\pacs{24.30-v, 25.45.-z, 25.45.Hi, 24.10.-i}

\maketitle

\section{Introduction}

Theory of nucleon transfer reaction  formulated in terms of the matrix element  containing the potential transition operator is based on the perturbation approach over the potential transition operator.  It can be formulated in two forms: post or prior. In the post (prior) form  the initial (final) scattering wave function is approximated by a simpler  channel wave function.  The distorted wave Born approximation  (DWBA), which is the simplest approach,
 is the first order perturbation theory over the potential transition operator (which is different in the post and prior forms)  sandwiched by the initial and final channel wave functions. These channel  wave functions are given by a product of   bound state wave function of the  the initial (final nuclei)  multiplied by a corresponding distorted wave.  The DWBA is based on the assumption that  probability of  direct reactions is so small that  they can be treated as direct transition from the initial to final channel without any coupling to the other channels, what is not always true.  A definite improvement  is continuum discretized coupled channels (CDCC) method applied for the analysis of the deuteron stripping. In CDCC, in addition to $d+A$ channel,  the three-body breakup channel $p+n+A$ is included.  However,  the CDCC has its own limitations.  The main one  is  related with the contribution of the rearrangement channels.  For example, for deuteron stripping these rearrangement channels are the proton or neutron bound to the target. 
Because rearrangement channels  are not orthogonal to  the initial $d+ A$ channel and to the breakup $p+n+ A$  channel, their accurate inclusion makes the problem very complicated  and the only legitimate solution is the Faddeev formalism \cite{Faddeev}, which allows one to treat consistently non-orthogonal channels without double counting.  However,  it is quite difficult to use the Faddeev formalism on a routine basis and  its application, owing to the technical problems with the Coulomb interaction, is limited only to light nuclei.  Hence  CDCC is still useful, but one needs to clearly understand the  shortcomings of the CDCC and one of them is the absence of the rearrangement channels in the asymptotic regions.  

In practical calculations it is assumed that in a limited region near the target the CDCC wave function reproduces  the three-body wave function reasonably well.  To calculate the stripping matrix element  the standard iteration procedure is used:  the CDCC wave function does not  have  rearrangement channels in the asymptotic region but can be used to calculate the reaction matrix element contributed by the final volume around the target. Here the question of the uniqueness  appears: how the solution of the CDCC equation would change if we add  the rearrangement channel wave function to the original CDCC wave function. For example, if we consider the deuteron stripping reaction $d + A \to p + F$, where $F=(n\,A)$,  what would happen if we use   $\Psi_{i}^{CDCC(+)} +  \varphi_{F}\,\chi_{pF}^{(+)}$ as the initial wave function rather than just the CDCC wave function $\Psi_{i}^{CDCC(+)}$, where $\varphi_{F}$ is the  $(nA)$ bound state wave function and $\chi_{pF}$  is the $p-F$ distorted wave. It was shown 
in \cite{austern87} that  the presence of two optical potentials $U_{pA}$ and $U_{nA}$ suppresses the contribution from two rearrangement channels,
$p+ (nA)$ and $n + (pA)$  resolving the uniqueness problem.  However, if only one  optical potential is present, then the issue of uniqueness should be checked.  To suppress  the rearrangement channels  truncation over angular momentum  is being used.  Then a sensitivity to the maximal orbital angular momentum of the relative motion of  $p$ and $n$ $\,\,l_{pn}^{max}$ should be checked.   

Despite the shortcomings of the CDCC approach it remains the best option unless the Faddeev equations are solved.  
In this  work we use the CDCC approach to develop the theory of the deuteron stripping to resonance states.  However,  instead of the standard formulation of the theory with  the matix element expressed in terms of the volume integral we develop here the theory of the deuteron  stripping based on  the surface-integral formalism \cite{kadyrov09,bray} and  CDCC approach.  The first such  attempt has been done  in a recent work \cite{muk2011}, where both DWBA and CDCC method were used to derive the deuteron stripping reaction amplitude populating  bound states and resonances.  However, the CDCC part was not complete because  the surface integral was extended to the region where the CDCC  method fails.  Here we present another  formulation of the theory of the  stripping to resonance states  using the surface integral formulation based on the CDCC approach in a finite region around the target, that is  in the region where CDCC should work. In \cite{muk2011} the surface integral in the post form for stripping to bound states was taken over variable ${\rm {\bf r}}_{nA}$ while the  volume integral over the second Jacobian variable ${\bm \rho}_{pF}$  was taken over the whole space.  Now in our new formulation  the matrix element  is expressed in terms of the surface integral over ${\bm \rho}_{pF}$  at some finite $\rho_{pF}$ determined by the transition operator,  while the volume integral over the second Jacobian variable ${\rm {\bf r}}_{nA}$ is taken over  the limited  volume space because of the presence of the bound state wave function $\varphi_{F}$.  We also use the prior form for the analysis of the stripping to resonance states. In this case the matrix element can be expressed in terms of  surface integral over ${\bm \rho}_{dA}$ 
taken at some  finite  radius determined by the transition operator and the bound state wave function $\varphi_{pn}$ and the volume integral over the second Jacobian variable ${\rm {\bf r}}_{pn}$, which  is  taken over the limited  volume because of the presence of $\varphi_{pn}$.  Thus even for stripping to resonance the matrix element is taken over the limited space where CDCC works. 

We explore different choices of the channel wave functions and, correspondingly, different transition operators. 
One of the main unsolved problems in the conventional  theory for the deuteron stripping reaction $d + A \to p + F$  is  the inconsistency in the treatment of the $n-A$ potential, which should be real  to support the final bound or resonance state $(n\,A)$ but complex to describe the initial $n-A$ scattering.  We show how this inconsistency leads to the appearance of the auxiliary term  when connecting the conventional volume matrix element with the surface integral form. 
We also present  
calculations  using the FRESCO code \cite{fresco}  for stripping to bound states and resonances.  
The main goal of this work is to present an advanced theory of the deuteron stripping to a resonance, which further leads to the three-body continuum in the final state. Such reactions can occur in broad interval of the deuteron incident energies.
Note that the deuteron stripping to resonance requires $2.224$  MeV, the deuteron binding energy, to break the deuteron and additional energy to excite a resonance state. Hence the $Q$ value of the reaction is negative. That is why we do not consider here  deuteron stripping  at sub-Coulomb energies with new interesting physics  \cite{bertulani2007}. Such reactions can be studied using the Faddeev formalism. 
The theory, which we present here,  is aimed to analyze the deuteron stripping reactions  from low energies near the Coulomb barrier up to the deuteron incident energies $E_{d} \sim 100$ MeV. 
\section{Three-body theory of deuteron stripping populating bound states in the surface-integral formalism}
\label{three-body1}

Let us consider the deuteron stripping to bound states 
\begin{equation}
d + A \to p + F,
\label{deutstrbndst1}
\end{equation}
where $F=(A\,n)$ is the bound state. 

The reaction amplitude can be calculated exactly in the three-body model using the Faddeev integral equations in the Alt-Grassberger-Sandhas (AGS)  form \cite{alt67,alt2002,deltuva2009,deltuva2009a,deltuva2009b,deltuva2012} but it neglects internal degrees of freedom of the target or can only account for a few \cite{alt2007,deltuva2013}.   Moreover, the formalism is limited to targets with not too large charges. Nowadays, deuteron stripping on heavy nuclei with atomic number $A \sim 100$  are the most important and urgent because they can provide a missing vital information about $(n,\,\gamma)\,$ s-  or r-processes in stellar evolution.  The generalized Faddeev approach, which explicitly includes target excitations and the Coulomb interaction for arbitrary charges, was developed \cite{muk2012} but no computer codes based on the formalism are yet available.  Besides, the  Faddeev formalism is too complicated for use on an everyday basis, especially by experimental groups. 

In the traditional approach the reaction amplitude is calculated  using the iteration procedure. in which the volume matrix element containing the exact scattering wave function (in the initial state-post form or in the final state-prior form) is approximated 
by the one, in which the exact scattering wave function is replaced by some model wave function.  This approximation is used
because  nowadays there are no tools to calculate the many-body scattering wave function accurately, especially in the asymptotic regions with many open channels. Moreover, should this asymptotic behavior be available there is no need to calculate the matrix elements because the amplitude of the asymptotic outgoing wave in the corresponding channel is the reaction amplitude for transition to  this channel.  The idea of the iteration procedure is that  the matrix element  containing the  scattering wave function, which is not accurate asymptotically,
is still suited to calculate the reaction amplitude, because this matrix element is contributed by a limited volume around the target where the model scattering wave function may be accurate enough.  

First we consider the surface integral formalism in a three-body-model, in which all three particles are structureless and all the interaction potentials between them are real.  After that we specifically consider deuteron stripping reactions extending the three-body model
what requires using  optical potentials. Different options  and ways how they affect the reaction amplitude will be discussed.

We start from the consideration of  reaction (\ref{deutstrbndst1}) in the three-body model $p + n + A$.  We introduce the Jacobian variables ${\rm {\bf r}}_{\alpha}$ and ${\rm {\bm \rho}}_{\alpha}$ commonlly used to describe three-body systems, where $\,{\rm {\bf r}}_{\alpha}\,$ is the radius-vector connecting the center of masses of particles $\beta$ and $\gamma$ while $\,{\rm {\bm \rho}}_{\alpha}\,$ is the radius-vector connecting the center of mass of particle $\alpha$ and the center of mass of the system $\beta+ \gamma$  \cite{alt93}.
We also need a hyperradius in the six-dimensional configuration space defined according to
$X_{\alpha}=({ {\mu_{\alpha}}\,r_{\alpha}^{2}/m + { M_{\alpha}}\,\rho_{\alpha}^{2}/m})^{1/2}$, where $\,m$ is the nucleon mass and $\mu_{\alpha}$ is the reduced mass of particles $\beta$ and  $\gamma$, $\,M_{\alpha}= m_{\alpha}\,m_{\beta\,\gamma}/M$ is the reduced mass of particle $\alpha$ and the bound system $(\beta\,\gamma)$, and $\,\,M$ is the total mass of the three-body system.
Let us introduce the asymptotic region $\Omega_{\alpha}$ corresponding to the case when two particles $\beta$ and $\gamma$ are close to each other while the third particle $\alpha$ is far away. In this region $r_{\alpha}/\rho_{\alpha} \to 0$ at $\rho_{\alpha} \to \infty$ \cite{alt93}.  Also, we denote by
$\Omega_{0}$ the asymptotic region where all three particles are far away (breakup channel), that is 
$\,r_{\alpha},\,\rho_{\alpha} \to \infty$ and $\,r_{\alpha}/\rho_{\alpha} \to const \not=0$. The asymptotic behavior of the three-body wave function of three charged particles in different asymptotic regions was discussed in \cite{alt93,muk96,kadyrov09,kad2003,kad2004,bray}. 
For the asymptotic behavior of the three-body wave function we have 
\begin{align} 
\Psi_{\alpha}^{(+)} = \Psi_{\alpha}^{(0)} - \sum\limits_{\nu}\,\frac{M_{\nu}}{2\,\pi}\,{\mathcal M}_{\nu\,\alpha}^{(as)}\,u^{(+)}({ \rho}_{\nu})\,\phi_{\nu} +  \Psi_{0}^{(+)}.
\label{asymptwf1}
\end{align}
Here $\Psi_{\alpha}^{(+)}$ is the scattering wave function with the incident wave in the initial channel $\alpha$.  The two-cluster channel $\alpha$ is defined as the channel $\alpha + (\beta\,\gamma)$, where the free particle carries the name of the channel. For reaction (\ref{deutstrbndst1}) the incident channel $\alpha$ is $\,d+A$, that is $\alpha= A$ and $\Psi_{\alpha}^{(+)} \equiv \Psi_{{\rm {\bf k}}_{dA}}^{(+)}$ is the $d+ A$ scattering wave function calculated in the three-body model $p+n +A$. $\,\,{\rm {\bf k}}_{ij}$ is the relative momentum of particles $i$ and $j$; $\,\,\Psi_{\alpha}^{(0)}$ is the incident wave in the entry channel $\alpha$. The sum over the final two-body channels $\nu$ contains the elastic and rearrangement channels, $\,{\mathcal M}_{\nu\,\alpha}^{(as)}\,$ is the reaction amplitude leading to the final two-body channel $\nu$; $\,\phi_{\alpha} = \varphi_{\beta\,\gamma}$ is the bound state wave function of the pair $(\beta\,\gamma)$ in the channel $\alpha$; for example, for the channel $\,A+ d\,$, $\,\,\phi_{\alpha} = \varphi_{pn}\,$ and for the channel $\,\beta=p + (n\,A)\,$,  $\phi_{\beta} = \varphi_{nA}$.
Also $\,\,u^{(+)}(\rho_{\nu})\,$ is the outgoing wave in the two-fragment channel $\,\nu$. 
It should be understood that each $\nu$-th asymptotic term dominates in its asymptotic region $\Omega_{\nu}$. In the case of reaction (\ref{deutstrbndst1}) under consideration, $\alpha=A,\,\,\beta=p,\,\,\gamma=n$. $\,\Psi_{0}^{(+)}$ is the asymptotic component of $\Psi_{\alpha}^{(+)}$ in the asymptotic region $\Omega_{0}$.
We remind again that in the three-body model $\,p + n + A\,$ the nucleus $A$ is a structureless constituent particle, that is all the channels related to the target excitation and target breakup are neglected. 

Eq. (\ref{asymptwf1}) is of fundamental importance because it provides a model-independent definition of the reaction amplitude ${\mathcal M}_{\nu\,\alpha}^{(as)}$  as the amplitude of the outgoing spherical wave in the final channel $\nu$ formed from the initial channel $\alpha$ for an arbitrary collision of composite nuclei. However, its practical implementation in the many-body case, except for three- and four-body systems, is hardly yet possible, because contemporary microscopic methods
fail to provide the correct asymptotic behavior. That is why, if we are not going to use the Faddeev or Faddeev-Yakubovsli \cite{yakubovsky} coupled equations, the conventional methods for the determination of reaction amplitudes is to calculate the volume matrix elements in the post or prior forms. 

Below we remind how to derive these matrix elements in the three-body model, which can be extended to a many-body system. Let us consider the three-body wave function $\Psi_{\alpha}^{(+)}$ containing the incident wave in the channel $\alpha + (\beta\,\gamma)$. 
It satisfies the Schr\"odinger equation
\begin{align}
\Big(E -  H_{\alpha} - {\overline K}_{\alpha} - {\overline V}_{\alpha}\Big)\,\Psi_{\alpha}^{(+)}=0.
\label{SchreqPsia1}
\end{align}
Here, $\,H_{\alpha}=  K_{\alpha} + V_{\alpha}$ is the Hamiltonian describing the relative motion of the system $\beta + \gamma$, $\,K_{\alpha}$ is the kinetic energy operator of the relative motion of $\beta$ and $\gamma$, $\,{\overline K}_{\alpha}\,$ is the kinetic energy operator of the relative motion of $\alpha$ and the center of mass of $\beta + \gamma$, $\,{\overline V}_{\alpha}= V- V_{\alpha}$, $\,V= V_{\alpha} + V_{\beta} + V_{\gamma}\,$ is the total interaction potential in the three-body system, $\,V_{\alpha}\,$ is the interaction potential between $\beta$ and $\gamma$, $E = {\overline E}_{\alpha} - \varepsilon_{\alpha}$ is the total energy of three-body system, $\,{\overline E}_{\alpha}\,$ is the relative kinetic energy of the particle $\alpha$ and the pair $\,(\beta\,\gamma)$, $\,\varepsilon_{\alpha}=
m_{\beta} + m_{\gamma} - m_{\beta\,\gamma}$ is the binding energy of the bound state $\,(\beta\,\gamma)$,
$\,m_{\alpha}$ is the mass of particle $\alpha$.

Equation (\ref{SchreqPsia1}) can be rewritten in the channel $\beta \not= \alpha$ representation as
\begin{align}
\Big(E - H_{\beta} -  {\overline K}_{\beta} - {\overline V}_{\beta}\Big)\,\Psi_{\alpha}^{(+)}=0.
\label{SchreqPsiabeta1}
\end{align}
Note that, according to Eq. (\ref{asymptwf1}), $\,\Psi_{\alpha}^{(+)}\,$ has the incident wave only in the channel $\alpha$. 
Now neglecting the Coulomb interaction for a while (this does not affect the final result) we introduce the channel wave function in the channel $\beta$
\begin{align}
\Phi_{\beta}^{(0)}= e^{i\,{\rm {\bf q}}_{\beta} \cdot {\rm {\bm \rho}}_{\beta}}\,\phi_{\beta},
\label{Phibeta01}
\end{align}
where ${\rm {\bf q}}_{\beta}$ is the relative momentum of particle $\beta$ and the bound state $(\alpha\,\gamma)$, that is, the momentum conjugated to the Jacobian coordinate ${\rm {\bm \rho}}_{\beta}$. 
Multiplying Eq. (\ref{SchreqPsiabeta1}) from the left by the channel wave function $\Phi_{\beta}^{(0)}$
we get
\begin{align}
<\Phi_{\beta}^{(0)}\Big| \Big(E - {\overrightarrow H}_{\beta} -  {\overrightarrow{\overline K}}_{\beta}\Big)\Big|\Psi_{\alpha}^{(+)}> = <\Phi_{\beta}^{(0)} \Big| {\overline V}_{\beta}\Big|\,\Psi_{\alpha}^{(+)}>.
\label{PrSchreqPsiabeta1}
\end{align}
Taking into account that 
\begin{align}
(E - H_{\beta} -  {\overline K}_{\beta})\,\Phi_{\beta}^{(0)}=0
\label{Hbeta1}
\end{align}
we can rewrite
\begin{align}
<\Phi_{\beta}^{(0)}\Big| \Big( {\overleftarrow{\overline K}}_{\beta} -  {\overrightarrow{\overline K}}_{\beta}\Big)\Big|\Psi_{\alpha}^{(+)}> = <\Phi_{\beta}^{(0)} \Big| {\overline V}_{\beta}\Big|\,\Psi_{\alpha}^{(+)}>\,=\, {\mathcal M}_{\beta\,\alpha}. 
\label{PrSchreqPsiabeta2}
\end{align}
Here we took into account that the left-hand side is the conventional reaction amplitude ${\mathcal M}_{\beta\,\alpha}=
 <\Phi_{\beta}^{(0)} \Big| {\overline V}_{\beta}\Big|\,\Psi_{\alpha}^{(+)}>$.
The operator ${\overrightarrow{\overline K}}_{\beta}$ ($ {\overleftarrow{\overline K}}_{\beta}$) acts to the right (left).
When deriving this equation we took into account that $H_{\beta}$ is Hermitian
if $(\alpha\gamma)$ is a bound state, that is 
\begin{align} 
&<\Phi_{\beta}^{(0)}\Big| \Big( {\overleftarrow H}_{\beta} -  {\overrightarrow H}_{\beta}\Big)\Big|\Psi_{\alpha}^{(+)}>          
= <\Phi_{\beta}^{(0)}\Big| \Big( {\overrightarrow H}_{\beta} -  {\overrightarrow H}_{\beta}\Big)\Big|\Psi_{\alpha}^{(+)}> =0. 
\label{aux1}
\end{align}
It follows from the fact that $V_{\beta}$ is a Hermitian operator.
Because $\Phi_{\beta}^{(0)}$ contains the bound state $(\alpha\,\gamma)$ we can take the integral over ${\rm {\bf r}}_{\beta}$ by parts twice transforming ${\overleftarrow K}_{\beta}$ into ${\overrightarrow K}_{\beta}$. Hence $H_{\beta} = K_{\beta} + V_{\beta}$ is also the Hermitian operator. This validates Eq. (\ref{aux1}). 

Now using the Green's theorem 
\begin{align}
L & = < f({\rm {\bf r}}) |{\overleftarrow K} - 
{\overrightarrow K}| g({\rm {\bf r}})>   \nonumber\\
&= - \frac{1}{2 \mu^{2}} \lim \limits_{r \to \infty} r^{2} \int {\rm d}{\rm {\bf {\hat r}}} 
\left [g({\rm {\bf r}}) \frac{\partial {f^{*}({\rm {\bf r}})}}{{\partial r}}  -  
 f^{*}({\rm {\bf r}}) \frac{\partial {g({\rm {\bf r}})}}{{\partial r}} \right ] 
\label{greentheorem1}
\end{align} 
we can transform the volume integral on the left-hand side of Eq. (\ref{PrSchreqPsiabeta2}) into a surface one in the subspace over ${\rm {\bm \rho}}_{\beta}$:
\begin{widetext}
\begin{align}
 < \Phi _\beta ^{(0)}|({{\overleftarrow {\overline K}} _\beta } - {{\overrightarrow {\overline K}}_\beta })| \Psi _{\alpha }^{(+)} > 
= & -\frac{1}{{2{\kern 1pt} M_\beta ^2}}\mathop {\lim }\limits_{{\rho_\beta } \to \infty } \rho_\beta ^2\,\int\,{\rm d}\,{\rm {\bf r}}_{\beta}\,\phi_{\beta}^{*}\,\int {\text{d}} {{\rm {\hat {\bm \rho}}}_{\beta}}{\kern 1pt} 
\nonumber \\ & \times
{\text{  \Big[}}\Psi_{\alpha}^{(+)}\,{\text{ }}\frac{{\partial\,e^{-i\,{\rm {\bf q}}_{\beta} \cdot {\rm {\bm \rho}}_{\beta}}}}{{\partial {\rho _\beta }}} - e^{-i\,{\rm {\bf q}}_{\beta} \cdot {\rm {\bm \rho
}}_{\beta}}\,\frac{{\partial \Psi_{\alpha}^{(+)}}}{{\partial {\rho _\beta }}}{\text{ \Big]}}\,.
\label{surfint1}
\end{align} 
\end {widetext}
Taking into account that the leading asymptotic term of  $\Psi_{\alpha}^{(+)}$ in $\Omega_{\beta}$ is (see Eq. (\ref{asymptwf1}))
\begin{align}
\Psi_{\alpha}^{(+)} \stackrel{\Omega_{\beta}}{\approx} -\frac{M_{\beta}}{2\,\pi}\,{\mathcal M}_{\beta\,\alpha}^{(as)}\,u^{(+)}({\rho}_{\beta})\,\phi_{\beta}
\label{asymptOmegabeta1}
\end{align}
and using the asymptotic equation \cite{messiah}
\begin{align}
{e^{i{\kern 1pt} {{\mathbf{q}}_\beta }\cdot{{\bm{\rho}}_\beta }}}\xrightarrow{{{\rho _\beta } \to \infty }}
& \frac{1}{{2\pi {q_\beta }{\rho _\beta }}}   \Big[{e^{i{q_\beta }{\rho _\beta }}}\delta ({\rm  {\hat {\bf q}}}_{\beta} -   {\rm {\hat {\bm \rho}}} _{\beta })\,\, 
- \,\,{e^{ - i{q_\beta }{\rho _\beta }}}\delta ({\rm {\bf {\hat q}}}_{\beta } + {\rm {\hat {\bm \rho}}} _{\beta })\Big],
\label{deltaasympt1}
\end{align}
and the normalization integral 
\begin{align}
\int\,d{\rm {\bf r}}_{\beta}\,|\phi_{\beta}|^{2} = 1
\label{normintbeta1}
\end{align}
we get 
\begin{align}
&<\Phi_{\beta}^{(0)}\Big| \Big( {\overleftarrow{\overline K}}_{\beta} -  {\overrightarrow{\overline K}}_{\beta}\Big)\Big|\Psi_{\alpha}^{(+)}> = {\mathcal M}_{\beta\,\alpha}^{(as)}.  
\label{surfint2} 
\end{align}         
Hence
\begin{align}
&{\mathcal M}_{\beta\,\alpha}^{(as)}\, = \,{\mathcal M}_{\beta\,\alpha} . 
\label{Mabexact11}
\end{align}
Thus we have proven that the conventional reaction amplitude ${\mathcal  M}_{\beta\,\alpha}$  given by the volume matrix 
element coincides with the amplitude ${\mathcal M}_{\beta\,\alpha}^{(as)}$  of the outgoing scattered wave in the channel $\beta$  with the incident wave  in the channel $\alpha$.
In the standard applications to decrease the transition operator one can subtract the final channel potential $U_{\beta}$ in the matrix 
element on the right-hand-side of Eq. (\ref{PrSchreqPsiabeta2}) what leads
to the final channel wave function $\Phi_{\beta}^{(-)}= \chi_{\beta}^{(-)}\,\phi_{\beta}$,  where $\,\,\chi_{\beta}^{(-)}$ is the distorted wave generated by the channel potential $\,U_{\beta}\,$ and describing the scattering of particle $\,\beta\,$ and the bound state $\,(\alpha\,\gamma)\,$. The channel potential is arbitrary and can be real or complex.  From the derivation it is clear that the matrix element doesn't depend on the choice of $\,U_{\beta}\,$ if  $\,\Psi_{\alpha}^{(+)}\,$ is  the exact three-body wave function.  Then we have 
\begin{align}
{\mathcal M}_{\beta\,\alpha}  = 
<\Phi_{\beta}^{(-)}\Big| \Big( {\overleftarrow{\overline K}}_{\beta} -  {\overrightarrow{\overline K}}_{\beta}\Big)\Big|\Psi_{\alpha}^{(+)}>       
= <\Phi_{\beta}^{(-)} \Big| {\overline V}_{\beta} - U_{\beta}\Big|\,\Psi_{\alpha}^{(+)}>.
\label{Mabexact11}
\end{align}

After introducing the distorted wave in the channel $\beta$ we can turn on the Coulomb interaction. 

Now let us discuss the lessons which we can learn from derivation of  Eq. (\ref{Mabexact11}). 
\begin{itemize}
\item[1.] This equation proves that indeed the volume matrix element, which is used in standard calculations of the reaction amplitude ${\mathcal M}_{\beta\,\alpha}$, is, in fact, the amplitude ${\mathcal M}_{\beta \alpha}^{(as)}$ of the leading asymptotic term of the exact three-body scattering wave function in the asymptotic domain $\Omega_{\beta}$. 
\item[2.] Equation (\ref{SchreqPsiabeta1}) is important for deriving Eq. (\ref{Mabexact11}). The former shows 
that the exact scattering wave function $\Psi_{\alpha}$ also satisfies the Schr\"odinger equation in the channel $\beta$ representation, that is, it has correct asymptotic behavior in the channel $\beta \not= \alpha$. The corresponding integral equation for $\Psi_{\alpha}$ will be homogeneous in the $\beta \not= \alpha$ channel. 
\item[3.] There is a clear advantage of using the volume matrix element rather than to calculate the amplitude of the asymptotic scattering wave function in the corresponding asymptotic domain. 
Because $\phi_{\beta} = \varphi_{\alpha\,\gamma}$ is the bound state wave function of the pair $(\alpha\,\gamma)$, the integration over the Jacobian coordinate $r_{\beta}$ is limited. The transition 
operator $\,{\overline V}_{\beta} - U_{\beta}\,$, where $\,{\overline V}_{\beta}= V_{\alpha} + V_{\gamma}\,$, $\,\,V_{\alpha} \equiv 
V_{\beta\,\gamma}\,$ and $\,V_{\gamma} \equiv V_{\alpha\,\beta}\,$, cuts the integration over the second Jacobian variable $\rho_{\beta}$ at some finite value. Hence, it is sufficient to know the scattering wave function $\Psi_{\alpha}^{(+)}$, developing from the initial state $\Psi^{(0)}_{\alpha}$, only in the constrained domain in the coordinate space $\{{\rm {\bf r}}_{\beta}, {\rm {\bm \rho}}_{\beta}\}$ around target nucleus $\alpha$. 
\end{itemize}

Let us introduce $\,R_{\beta}\,$ as a quantity larger than the nuclear interaction radius $\,R_{\beta}^{N}\,$ in the two-body subsystem $\,(\alpha \gamma)\,$ and $\,{\overline R}_{\beta}\,$ to be a quantity larger than nuclear interaction radius in the two-cluster channel $\beta$. These are the values which should be taken into account in the volume matrix element to achieve the required accuracy, which is typically $\sim 1\%$ or better.

 It is worth mentioning that $\,R_{\beta}\,$ may be taken significantly larger than the nuclear interaction radius $\,R^{N}_{\beta}\,$ of particles $\,\alpha\,$ and $\,\gamma\,$. Clearly $\,R_{\beta}\,$ should be larger than  $\,1/\kappa_{\beta}\,$, where $\,\kappa_{\beta} \equiv \kappa_{\alpha\,\gamma}\,$ is the bound state wave number  of the bound state $\,(\alpha \gamma)$. We define a hyper-radius corresponding to $\,\{R_{\beta}, \,{\overline R}_{\beta}\}\,$ as $X_0= ({ {\mu_{\beta}}\,R_{\beta}^{2}/m + { M_{\beta}}\,{\overline R}_{\beta}^{2}/m})^{1/2}$.
With this we can rewrite  Eq. (\ref{Mabexact11}) as
\begin{widetext}
\begin{align}
{\mathcal M}_{\beta\,\alpha} & \approx  <\Phi_{\beta}^{(-)} \Big| {\overline V}_{\beta} - U_{\beta}\Big|\,\Psi_{\alpha}^{(+)}>\Big|_{X_{\beta} \leq { X}_{0}}          
 =<\Phi_{\beta}^{(-)}\Big| \Big( {\overleftarrow{\overline K}}_{\beta} -  {\overrightarrow{\overline K}}_{\beta}\Big)\Big|\Psi_{\alpha}^{(+)}>\Big|_{X_{\beta} \leq { X}_{0}}         \nonumber\\
&=- \frac{{\overline R}_{\beta}^{2}}{2\,M_{\beta}}\,\int\limits_{r_{\beta} \leq  R_{\beta}}\,{\rm d}{{\rm {\bf r}}_{\beta}}\,\phi_{\beta}^{*}\,\int\,{\rm d}\,\Omega_{{\rm {\bm { \rho}}}_{\beta}}\,\Big[\Psi_{\alpha}^{(+)}\,\frac{{\partial \chi_{\beta}^{(-)*}}}{{\partial {\rho _\beta}}}      - \chi_{\beta}^{(-)*}\,\frac{{\partial \Psi_{\alpha}^{(+)}}}{{\partial {\rho _\beta }}}\Big]\Big|_{\rho_{\beta}= {\overline R}_{\beta}}. 
\label{rhobetaM1}
\end{align}
\end{widetext}

Eq. (\ref{rhobetaM1}) is our first main result in this section. It shows that in the three-body method  the volume matrix element can be transformed into the peripheral matrix element.  The surface integral over $\,\Omega_{{\rm {\bm { \rho}}}_{\beta}}\,$ in Eq. (\ref{rhobetaM1}) is taken along the sphere  with the radius $\,\rho_{\beta}={\overline R}_{\beta}\,$ encircling the finite volume inside of this sphere, while the integral over $\,{\rm {\bf r}}_{\beta}\,$ is taken over the volume confined by the sphere with the radius $\,r_{\beta} = R_{\beta}$.  If we take the limit ${\overline R}_{\beta} \to \infty$  we get identitty, $\,{\mathcal M}_{\beta\,\alpha} \equiv  {\mathcal M}_{\beta\,\alpha}^{(as)}$.  However, in practical calculations we can constrain the integration region by a fintie $\,{\overline R}_{\beta}$, that is, we need to know the wave function $\,\Psi_{\alpha}^{(+)}\,$ only in a limited volume around the target.  The value of  $\,{\overline R}_{\beta}\,$  can be determined by checking the convergence of the matrix element as function of $\,{\overline R}_{\beta}$.   If  $\,{\overline R}_{\beta}\,$  is not too large then we do not need to know the asymptotic behavior  of  $\,\Psi_{\alpha}^{(+)}$.
Although Eq. (\ref{rhobetaM1}) has been derived in a three-body model, the derivation is valid also for a many-body case assuming  that $\,\Psi_{\alpha}^{(+)}\,$ is the exact many-body scattering wave function with the incident wave in the channel $\alpha + (\beta\gamma)$.  

Note that in the prior formalism the stripping reaction matrix element is given by
\begin{widetext}
\begin{align}
{\mathcal M}_{\beta\,\alpha} & \approx  <\Psi_{\beta}^{(-)} \Big| {\overline V}_{\alpha} - U_{\alpha}\Big|\,\Phi_{\alpha}^{(+)}>\Big|_{X_{\alpha} \leq { X}_{0}}          
 =<\Psi_{\beta}^{(-)}\Big| \Big( {\overleftarrow{\overline K}}_{\alpha} -  {\overrightarrow{\overline K}}_{\alpha}\Big)\Big|\Phi_{\alpha}^{(+)}>\Big|_{X_{\alpha} \leq { X}_{0}}         \nonumber\\
&=-\frac{{\overline R}_{\alpha}^{2}}{2\,M_{\alpha}}\,\int\limits_{r_{\alpha} \leq  R_{\alpha}}\,{\rm d}{{\rm {\bf r}}_{\alpha}}\,\phi_{\alpha}\,\int\,{\rm d}\,\Omega_{{\rm {\bm { \rho}}}_{\alpha}}\,\Big[\chi_{\alpha}^{(+)}\,\frac{{\partial \Psi_{\beta}^{(-)*} }}{{\partial {\rho _\alpha}}}     -  \Psi_{\beta}^{(-)*}\,\frac{{\partial \chi_{\alpha}^{(+)}}}{{\partial {\rho _\alpha }}}\Big]\Big|_{\rho_{\alpha}= {\overline R}_{\alpha}},
\label{rhoalphaM1}
\end{align}
\end{widetext}
where $\phi_{\alpha} = \varphi_{\beta\,\gamma}$.
Though the  post and prior forms are identical,  there are  computational advantages in using specific form depending on the reaction under consideration. 
We will address it below. 

The fact that the integration volume is constrained  is quite important because it justifies the usage of the different approximations for the exact scattering wave function, which are valid in the limited space around nucleus even if these approximations do not provide wave functions with correct asymptotic behavior in the rearrangement channels. 
Such approximations are well known: distorted wave Born approximation (DWBA), continuum discretized coupled 
channels (CDCC) \cite{austern87,austern88,austern89,austern96} and adiabatic method (ADWA) \cite{johnson70}. In the DWBA the initial scattering wave function contains only the contribution from the incident channel $\alpha + (\beta\,\gamma)$. In the CDCC method the initial wave function is contributed by the channel $\alpha + (\beta\,\gamma)$, in which the pair $(\beta\,\gamma)$ is taken in the bound state plus discretized  states describing the three-body system $\alpha  + \beta + \gamma $ in the continuum.  The adiabatic approach, as the CDCC, also takes into account the continuum states of the $(\beta\,\gamma)$ system but in a more simplified way. All three methods fail to provide correct asymptotic behavior in the rearrangement channels.
Nevertheless, all three methods, being not perfect, still give reasonable transfer reaction cross sections. The accuracy of the each  method depends on the kinematics, energy, interacting nuclei and purposes. When the energy increases the contribution of the deuteron breakup channel also increases making the ADWA and CDCC more adequate than the DWBA.   In addition, this creates another problem to be dealt with: it is the increase of the contribution from the nuclear interior. In the internal region a strong coupling of different channels occurs and antisymmetrization effects are important.  Meantime the existing approaches, DWBA ,  ADWA and CDCC,  are based on the three-body model extended by adopting optical potentials and they are designed to treat mostly peripheral reactions.  The surface-integral formalism developed here in the combination with the $R$-matrix method can provide a solution.

Finally one important feature of Eq. (\ref{rhobetaM1}) remains to be discussed. Assume that we use the CDCC wave function to calculate $\Psi_{\alpha}^{(+)}$. In the CDCC method particles $\beta$ and $\gamma$ are kept close to each other by using the projection operator, which truncates the number of the allowed $\beta-\gamma$ partial waves. At the same time the surface integral over $\Omega_{{\rm {\bm { \rho}}}_{\beta}}$ is calculated at $\,\rho_{\beta}={\overline R}_{\beta}\,$. As $\,{\overline R}_{\beta}\,$ can be significantly larger than the nucleus radius, the dominant contribution to the volume integral over $\,r_{\beta}\,$
should come from $\,R_{\beta}^{N} \leq r_{\beta} \leq R_{\beta}$. Hence the reaction amplitude given by Eq. (\ref{rhobetaM1}) is entirely peripheral in the subspace over $\,{\rm {\bf r}}_{\beta}\,$ and  $\,{\bm \rho}_{\beta}\,$   and can be rewritten as        
\begin{widetext}
\begin{align}
&{\mathcal M}_{\beta\,\alpha} 
=- \frac{{\rho_0}^{2}}{2\,M_{\beta}}\,\int\limits_{R_{\beta}^{N} \leq r_{\beta} \leq   R_{\beta}}\,{\rm d}{{\rm {\bf r}}_{\beta}}\,\phi_{\beta}^{*}\,\int\,{\rm d}\,\Omega_{{\rm {\bm { \rho}}}_{\beta}}\,\Big[\Psi_{\alpha}^{CDCC(+)}\,\frac{{\partial \chi_{\beta}^{(-)*}}}{{\partial {\rho _\beta}}}  - \chi_{\beta}^{(-)*}\,\frac{{\partial \Psi_{\alpha}^{CDCC(+)}}}{{\partial {\rho _\beta }}}\Big]\Big|_{\rho_{\beta}= {\overline R}_{\beta}}, 
\label{rhobetaM2}
\end{align}
\end{widetext}
where  $\phi_{\beta}(r_{\beta}) \approx C_{\beta}\,W_{-\eta_{\beta},l_{\beta}+1/2}(2\,\kappa_{\beta}\,r_{\beta})/r_{\beta}$ is the radial part of the bound state wave function, $\,\,C_{\beta}$ is the asymptotic normalization 
coefficient   (ANC) of the bound state $(\alpha\gamma)$, $W_{-\eta_{\beta},l_{\beta}+1/2}(2\,\kappa_{\beta}\,r_{\beta})$ is the Whittaker function, $\eta_{\beta}$ is the Coulomb parameter and $l_{\beta} \equiv l_{\alpha\gamma}$ is the orbital angular momentum of the bound state $(\alpha\gamma)$. In the many-body case $\phi_{\beta}$ should be replaced by the corresponding overlap function. 
Transition from the three-body model to the CDCC requires using of the optical potentials, which effectively take  into account the internal structure
of the target.  	


\section{Deuteron stripping to a bound state. From many-body to three-body model}
\label{deutstripbstpost1}

\subsection{Post form}

In the previous section we considered  the deuteron stripping reaction in the three-body problem, that is all three particles, $p,\,n$ and $A$ are structureless constituents.  Hence all the interaction potentials are real.  Definitely internal degrees of freedom of  the target
should be taken into account. However, a rigorous practical many-body theory of transfer reactions is not yet available and  contemporary nuclear reaction theory uses the three-body model in which the internal structure of the target is taken  into account effectively by replacing $N-A$ optical potentials.  

Here we consider this reduction of the many-body problem to the three-body one  and apply the surface-integral formalism 
developed in the previous section  specifically for the deuteron stripping reaction. We neglect the antisymmetrization between the existing proton and the rest of the nucleons in the target $A$.  To derive an equation for the reaction amplitude we start from the Schr\"odinger equation for the total scattering wave function $\,\Psi_{i}^{(+)}\,$ developing from the initial channel: 
\begin{align}
\big(E- K_{pF} - K_{nA}- V_{nA}- V_{pA}- V_{pn}- H_{A}\big)\,\Psi_{i}^{(+)}=\,0,
\label{SchreqPsii1}
\end{align}
where $\,V_{nA}\,$ ($V_{pA}$) is the $\,n-A\,$ ($p-A$) interaction
potential given by the sum of $\,NN\,$ potentials (three-body forces can also be included), $\,H_{A}$ is the internal Hamiltonian of nucleus $A$.  $\,\,\Psi_{i}^{(+)}$ has the incident wave in the initial channel $d + A$ and outgoing waves in both direct and rearrangement channels.  

\subsubsection{Standard choice of the exit channel wave function}

To proceed further we need to adopt a suitable form of the final channel wave function.
Here we show how to derive and transform the stripping reaction amplitude in the case when the exit channel wave function is taken in the following standard form
\begin{align}
\Phi_{f}^{(-)}= \chi_{pF}^{(-)}\,\varphi_{F},
\label{finchwf1}
\end{align}
where $\chi_{pF}^{(-)}$ is the distorted wave of particles $p$ and $F$ in the final channel and  $\,\varphi_{F}$ is the bound state wave function of nucleus $F$ in the final channel.  The wave function $\Phi_{f}^{(-)}$ is a solution of the Schr\"odinger equation
\begin{align}
\big(E - K_{pF}- U_{pF} - K_{nA}- V_{nA} - H_{A}\big)\,\Phi_{f}^{(-)}=0.
\label{SchreqnPhif1}
\end{align}

Multiplying Eq. (\ref{SchreqPsii1}) from the left by $\Phi_{f}^{(-)*}$ and taking into account Eq. (\ref{SchreqnPhif1}) we get 
\begin{widetext}
\begin{align}
< & \Phi_{f}^{(-)} \big|E- {\overrightarrow K}_{pF} - {\overrightarrow K}_{nA} - V_{nA} - {\overrightarrow H}_{A}- [V_{pA} + V_{pn} - U_{pF}] - U_{pF}\big| \Psi_{i}^{(+)} >                \nonumber\\ 
&=  <\Phi_{f}^{(-)} \big| {\overleftarrow K}_{pF} - {\overrightarrow K}_{pF}  +  {\overleftarrow K}_{nA} - {\overrightarrow K}_{nA} +  {\overleftarrow H}_{A} - {\overrightarrow H}_{A}- [V_{pA} + V_{pn} - U_{pF}]   \big| \Psi_{i}^{(+)}>                         \nonumber\\
&= <\Phi_{f}^{(-)} \big|{\overleftarrow K}_{pF} - {\overrightarrow K}_{pF} - [V_{pA} + V_{pn} - U_{pF}] \big| \Psi_{i}^{(+)}>=0.
\label{Schreqprfnch2}
\end{align}
\end{widetext}
When deriving this equation we took into account that the operators 
$H_{A}$ and $K_{nA} $ are Hermitian because the final channel wave function contains the bound state $F=(nA)$.  Hence,  $<\Phi_{f}^{(-)} \big|{\overleftarrow K}_{nA}  + {\overleftarrow H}_{A} - {\overrightarrow K}_{nA} - {\overrightarrow H}_{A} \big| \Psi_{i}^{(+)}> \,=0$.
We can rewrite Eq. (\ref{Schreqprfnch2}) as
\begin{align}                         
&{\mathcal M}^{(as)}= <\Phi_{f}^{(-)} \big|{\overleftarrow K}_{pF} - {\overrightarrow K}_{pF}\big| \Psi_{i}^{(+)}>      
\label{kinenmatrelpost1}\\
&=<\Phi_{f}^{(-)} \big|V_{pA} + V_{pn} - U_{pF}\big| \Psi_{i}^{(+)}> \equiv {\mathcal M}^{(post)} .
\label{reactamplvol1}
\end{align}
We can verify that the matrix element  $<\Phi_{f}^{(-)} \big|{\overleftarrow K}_{pF} - {\overrightarrow K}_{pF}\big| \Psi_{i}^{(+)}> $  is equal to the amplitude $\,{\mathcal M}^{(as)}$ of the leading  asymptotic term of the exact $d+A$ scattering wave function $\Psi_{i}^{(+)}$  in the channel $p+F$. It can be proved  by converting matrix element (\ref{kinenmatrelpost1}) into a surface integral in the subspace over ${\rm {\bf \rho}}_{pF}$. 
After taking the limit of the radius of the surface $\rho_{pF} \to \infty$ we get that the matrix element is nothing but the reaction amplitude $\,{\mathcal M}^{(as)}$ \cite{kadyrov09,bray}. 
This amplitude is the model-independent definition of the reaction amplitude.  
Thus it follows from  Eq. (\ref{reactamplvol1})  that  the conventional reaction amplitude given by the volume matrix element ${\mathcal M}^{post}$  is equal to   $\,{\mathcal M}^{(as)}\,$ .
In Eq. (\ref{reactamplvol1})  the  internal degrees of freedom of the target $A$ are taken  into account properly.  However, the exact many-body scattering wave function is not yet available and at this stage approximations are supposed to be used.  

First we use the fact that, owing to the presence of the factor $\varphi_{F}\,[V_{pA} + V_{pn} - U_{pF}]$,  the integration can be carried over a finite volume in the 6-dimensional configuration space $\{{\rm {\bm \rho}}_{pF},\, {\rm {\bf r}}_{nA}\}$, where we do not need to know the asymptotic behavior of the scattering wave function $\Psi_{i}^{(+)}$.  The presence of the factor  $\varphi_{F}\,[ V_{pA} + V_{pn} - U_{pF}]$  in the matrix element  constrains the integration over the Jacobian variables by a finite volume around the target nucleus.  Clearly $\varphi_{F}$ cuts the integration over the internal nucleon coordinates  including the coordinates of the transferred neutron.  We introduce ${\mathcal R}_{nA} $ as  the maximal $r_{nA}$, which is required to achieve a desired accuracy for the integral over $r_{nA}$. We also introduce ${\mathcal R}_{pF} $ as  the maximal $\rho_{pF}$, which is required to achieve a desired accuracy for the integral over $\rho_{pF}$.  If $\,R_{nA}\,$ is the channel radius for which we can use the radius of the strong $n-A$ interaction, then $\,{\mathcal R}_{nA} > R_{nA}\,$ and may be significantly larger for loosely bound states.  At some large enough $\,\rho_{pF}\,$ and finite  $\,r_{nA} \leq {\mathcal R}_{nA}\,$  the nuclear part  $ V_{pA}^{N} + V_{pn} - U_{pF}^{N}$  of the transition operator becomes negligible. 

Now we consider the matrix element $\,<\chi_{f}^{(-)}\,\varphi_{F} \big| V_{pA}^{C}  - U_{pF}^{C}\big|\Psi_{i}^{(+)}>\,$  from the Coulomb part of the transition operator.  At $\,r_{pA} >> R_{A}$, where $\,R_{A}\,$ is the radius of nucleus $\,A$, we can approximate in the leading order $V_{pA}^{C}(r_{pA}) \approx U_{pA}^{C}(r_{pA}) \,=\, Z_{A}\,e^{2}/r_{pA}$ while $U_{pF}^{C}(\rho_{pF})= Z_{A}\,e^{2}/\rho_{pF}$, where $Z_A$ is the charge of nucleus $A$.
Taking into account that
\begin{align}
{\rm {\bm \rho}}_{PF}= {\rm {\bf r}}_{pA} -  \frac{1}{A+1}\,{\rm {\bf r}}_{nA},
\label{rhopFrpA1}
\end{align}
we get for $\,r_{pA} >> r_{nA}$ 
\begin{align}
U_{pF}^{C}(\rho_{pF}) - U_{pA}(r_{pA}) \stackrel{r_{pA} >> r_{nA}}{\approx}   
\frac{Z_{A}\,e^{2}}{r_{pA}}\,\frac{1}{A+1}\,\frac{{\rm {\bf {\hat r}}}_{pA} \cdot{\rm {\bf r}}_{nA}}{r_{pA}},
\label{Coulpot1}
\end{align}
where ${\rm {\bf {\hat r}}}= {\rm {\bf r}}/r$ and $A$ also represents the total number of nucleons in nucleus $A$. Hence, at large enough $r_{pA}$  the difference in the Coulomb potential becomes negligible, that is, the integration volume in the matrix element $\,<\chi_{f}^{(-)}\,\varphi_{F} \big|\, U_{pA}^{C}  - U_{pF}^{C}\,\big|\Psi_{i}^{(+)}>\,$ is also limited. 
Then we can rewrite 
\begin{align}                         
{\mathcal M}^{(post)} & = <\Phi_{f}^{(-)} \big|V_{pA} + V_{pn} - U_{pF}\big| \Psi_{i}^{(+)}> \Big|_{X \leq X_0}   
\label{volmatrel1}  \\
&=\,<\Phi_{f}^{(-)} \big|{\overleftarrow K}_{pF} - {\overrightarrow K}_{pF}\big| \Psi_{i}^{(+)}>\Big|_{X \leq X_0} ,
\label{reactamplapp1}
\end{align}
where the hyperradius is defined as
\begin{align}
X= \sqrt{\frac{\mu_{nA}}{m}\,r_{nA}^{2} + \frac{\mu_{pF}}{m}\,\rho_{pF}^{2}}
\label{hyperradiusX1}
\end{align}
and
\begin{align}
X_0= \sqrt{\frac{\mu_{nA}}{m}\,{\mathcal R}_{nA}^{2} + \frac{\mu_{pF}}{m}\,{\mathcal R}_{pF}^{2}},
\label{hyperradiusR1}
\end{align}
$m$ is the nucleon mass,  $\mu_{ij}$ is the reduced mass of particles $i$ and $j$.  

Transforming now the matrix element containing the kinetic energy operators into a surface integral in the subspace over ${\rm {\bf \rho}}_{pF}$ we get 
\begin{widetext}
\begin{align}                         
{\mathcal M}^{(post)}=& 
\,- \frac{{\mathcal R}_{pF}^{2}}{2\,\mu_{pF}} \int\,{\rm d}\,{ \zeta}_{F}\,\varphi_{F}^{*}(\zeta_{F})\,\int{\rm d}\,\Omega_{{\rm {\bm \rho}}_{pF}}\,
\nonumber \\ & \times
\big[\Psi_{i}^{(+)}\,\frac{
\partial\,\chi_{pF}^{(-)*}\,({\rm {\bm  \rho}}_{pF})}{\partial\,\rho_{pF}} - \chi_{pF}^{(-)*}({\rm {\bm  \rho}}_{pF})\,\frac{
\partial\,\Psi_{i}^{(+)}}{\partial\,\rho_{pF}} \big]\Big|_{\rho_{pF}={\mathcal R}_{pF}; \,r_{nA} \leq {\mathcal R}_{nA}} .
\label{reactamplsurf2}
\end{align}
\end{widetext}
Here the surface integral is taken over the sphere with the radius $\rho_{pF}= {\mathcal R}_{pF}$ while the volume integral is taken over the set $\zeta_{F}$ of the internal coordinates of nucleus $F$  subject to a condition that the coordinate $r_{nA}$ is constrained by $r_{nA}  \leq {\mathcal R}_{nA}$.  Thus the stripping matrix element is contributed by the finite volume in the space $\{{\rm {\bm \rho}}_{pF},\,{\rm{\bf r}}_{nA}\}$.   This important fact paves the way for different approximations used in the contemporary nuclear reaction theory, because  within this finite volume the exact initial scattering wave function $\Psi_{i}^{(+)}$  can be approximated by wave functions, which do not have correct asymptotic behavior in the rearrangement channel $p+F$. Nevertheless, they approximate this wave function in the finite volume fairly enough, at least in the three-body approach.  Such approximations are well known:  the initial channel wave function $\chi_{dA}^{(+)}\,\varphi_{pn}\,\varphi_{A}$ used in the DWBA,  the CDCC  wave function $\Psi_{i}^{CDCC(+)}\,\varphi_{A}$ or the adiabatic  model wave function $\Psi_{i}^{AD(+)}\,\varphi_{A}$. 
Note that all the three approaches are based on the three-body model, in which the target $A$ is treated
as structureless constituent particle. That is why in each approach the scattering wave function contains the target bound state wave function $\varphi_{A}$ in a factorized form.  The composite structure of the target is taken into account  effectively via the optical potentials.

Eq. (\ref{volmatrel1})  is exact if the antisymmetrization effects are neglected.  Assume now that in the integration region the wave function  $\Psi_{i}^{(+)}$  can be approximated  by the wave functions used in the DWBA, CDCC \cite{austern87} or ADWA \cite{johnson70}. Usually such an approximation is done in the volume matrix element  (\ref{volmatrel1}). Here we apply it after transforming  the volume matrix element into the surface integral  over ${\rm {\bf \rho}}_{pF}$  keeping the volume integral over $\zeta_{F}$.  This is the main difference between the standard approach and the one we use here. 
The replacement of $\Psi_{i}^{(+)}$ by  the CDCC wave function, which is the most advanced among the three above mentioned methods,  leads to  the  following CDCC reaction amplitude in the surface approximation:
\begin{widetext}
\begin{align}                         
{\mathcal M}_{surf}^{CDCC(post)} = & \,- \frac{{\mathcal R}_{pF}^{2}}{2\,\mu_{pF}} \int\,{\rm d}\,{\rm {\bf  r }}_{nA}\,I_{A}^{F*}({\rm {\bf r}}_{nA})\,\int{\rm d}\,\Omega_{{\rm {\bm \rho}}_{pF}}\,
\nonumber \\ & \times
\big[\Psi_{i}^{CDCC(+)}\,\frac{
\partial\,\chi_{pF}^{(-)*}({\rm {\bm  \rho}}_{pF})}{\partial\,\rho_{pF}} - \chi_{pF}^{(-)*}({\rm {\bm  \rho}}_{pF})\,\frac{
\partial\,\Psi_{i}^{CDCC(+)}}{\partial\,\rho_{pF}} \big]\Big|_{\rho_{pF}={\mathcal R}_{pF}; \, r_{nA} \leq {\mathcal R}_{nA}} .
\label{CDCCsurf1}
\end{align}
\end{widetext}
Here $I_{A}^{F}({\rm {\bf r}}_{nA})= <\varphi_{A}\big|\varphi_{F}>$  is the overlap function of the bound state wave functions of nuclei $F$ and $A$. We remind that here we neglected  antisymmetrization effects. Thus starting from the exact volume matrix element, we transformed it into the surface integral over ${\rm {  \rho}}_{pF}$  leaving the volume integral over the second Jacobian variable ${\rm {\bf r}}_{nA}$.  After that the exact scattering wave function was replaced by the CDCC one  reducing the exact amplitude ${\mathcal M}^{(post)}$ in the surface-integral representation to the CDCC amplitude ${\mathcal M}_{surf}^{CDCC(post)}$ also in the surface-integral form. 

Now the question is  how this ${\mathcal M}_{surf}^{CDCC(post)}$ amplitude in the surface-integral form is related to the conventional CDCC 
amplitude given by the volume matrix element? Note that the conventional CDCC amplitude
\begin{align}
{\mathcal M}_{conv}^{CDCC(post)}= <\chi_{pF}^{(-)}\,I_{A}^{F}\, \big|U_{pA} + V_{pn} - U_{pF}\big| \Psi_{i}^{CDCC(+)}> \Big|_{X \leq X_0} 
\label{CDCCpostconv1}
\end{align}
is also obtained  from the exact matrix element  (\ref{reactamplvol1})  by  approximating
$\Psi_{i}^{(+)} \to \Psi_{i}^{CDCC(+)}$  and $\,V_{PA} \to U_{pA}$.  
To answer this question we transform Eq. (\ref{CDCCsurf1}) back to the volume integral. To do it we replace the surface integral by the volume integral in which the transition operator is given by the difference of the kinetic energy
operators ${\overleftarrow K} - {\overrightarrow K}$:
\begin{widetext}
\begin{align}                         
{\mathcal M}_{surf}^{CDCC(post)} &= <\chi_{pF}^{(-)}\,I_{A}^{F} \big|{\overleftarrow K}_{pF} - {\overrightarrow K}_{pF}\big| \Psi_{i}^{CDCC(+)}>\Big|_{X \leq X_0}  
\label{reactamplCDCC21}  \\
&=   <\chi_{pF}^{(-)}\,I_{A}^{F} \big|{\overleftarrow K} - {\overrightarrow K}\big| \Psi_{i}^{CDCC(+)}>\Big|_{X \leq X_0}  
\label{reactamplCDCC22}          \\
& = <\chi_{pF}^{(-)}\,I_{A}^{F} \big|U_{pA} + U_{nA} +V_{pn} -  V_{nA}^{sp} - U_{pF}  \big| \Psi_{i}^{CDCC(+)}>\Big|_{X \leq X_0}. 
\label{reactamplCDCCf2}
\end{align}
\end{widetext}
Here to get Eq. (\ref{reactamplCDCC22})  from Eq. (\ref{reactamplCDCC21}) we took into account that the matrix element ${\overleftarrow K}_{nA} - {\overrightarrow K}_{nA}$ vanishes because, after two integrations by parts over  $r_{nA}$,  the surface integral at $r_{nA} \to \infty$ disappears owing to the presence of the overlap function  $I_{A}^{F}$, and  ${\overleftarrow K}_{nA}$ can be converted into ${\overrightarrow K}_{nA}$.  Note that although the  integration over $r_{nA}$  is restricted by $r_{nA} \leq  {\mathcal R}_{nA}$, we can extend it to  infinity  to make the matrix element  from ${\overleftarrow K}_{nA} - {\overrightarrow K}_{nA}$ vanish. 
To get Eq. (\ref{reactamplCDCCf2})   we took into account that the  CDCC wave function  is  the solution of the Schr\"odinger equation 
\begin{align}
\big(E - T - U_{pA} - U_{nA} -  V_{pn}\big)\,\Psi_{i}^{CDCC(+)}
=0. 
\label{CDCC1}
\end{align}
Note that often the truncation of the relative orbital angular momentum $l_{pn}$ is used  in the CDCC approach \cite{austern87}, which works as an additional   suppression of the rearrangement channels \cite{austern88}  to the optical potentials $U_{pA}$ and $U_{nA}$.  This truncation is achieved by using the projector
\begin{align}
{\hat P}_{pn}= \sum\limits_{l_{pn}=0}^{l_{pn}^{max}}\,\sum\limits_{m_{l_{pn}}=-l_{pn}}^{l_{pn}}\,\int\,{\rm d}\Omega_{{\rm {\bf r}}_{pn}}\,
Y_{l_{pn}\,m_{l_{pn}}}({\rm {\bf{\hat r}}}_{pn})\,Y_{l_{pn}\,m_{l_{pn}}}^{*}({\rm {\bf {\hat r}}}_{pn}').
\label{Ppnprojector1}
\end{align}
Suppression of the rearrangement channels is required to provide a unique solution  of the CDCC  Schr\"odinger equation  (\ref{CDCC1}). 
It has been shown  in \cite{austern88}  that the suppression of the rearrangement channels  by the optical potentials is stronger 
than by the projection operator  ${\hat P}_{pn}$ and, a priori, there is no need to introduce the projector  ${\hat P}_{pn}$  
if  two optical potentials  $U_{pA}$ and  $U_{nA}$ are being used. But the constrain over $l_{pn}$  is always can be added if needed. 

We have assumed also that the overlap function $I_{A}^{F}$ is proportional  to the single-particle bound state wave function at all $r_{nA}$. Then $\chi_{pF}^{(-)}\,I_{A}^{F}$  satisfies the Schr\"odingier equation
\begin{align}
\big(E - K - V_{nA}^{sp}- U_{pF}\big)\,\chi_{pF}^{(-)}\,I_{A}^{F}=0,
\label{shreqnPhif21}
\end{align}
where $V_{nA}^{sp}= <\varphi_{A}|V_{nA}|\varphi_{A}>$  is the single-particle $n-A$ potential supporting the bound state.
 
There is an important point to be discussed here.  The integration  in Eq. (\ref{reactamplCDCCf2})  is taken at fixed $\rho_{pF} = {\mathcal R}_{pF}$ and $r_{nA} \leq {\mathcal R}_{nA}$, meaning that  the integration over $r_{pA}$  is also constrained.  These constraints follow from the ones  in the original matrix element  (\ref{reactamplsurf2}).  Replacing the exact scattering wave function 
$\Psi_{i}^{(+)} $ by $ \Psi_{i}^{CDCC(+)}\,\varphi_{A}$ in (\ref{reactamplsurf2})  we still  keep the constraints of the integration region as in the original matrix element. This is  because the CDCC method is valid  only in the limited  hypervolume with $X  \leq  X_0$,  where the asymptotic regime of $\Psi_{i}^{(+)}$ has not yet been reached. Within this volume the CDCC wave function is supposed to be a reasonable approximation to the exact one.   

Note that the transition operator in Eq. (\ref{reactamplCDCCf2})  differs from the one in the conventional CDCC amplitude (\ref{CDCCpostconv1})  and  the difference is due to the additional transition operator $U_{nA} - V_{nA}^{sp}$.  The appearance of this additional  transition operator is the price we pay for using energy-independent potentials. Because of importance of this issue we would like to trace the appearance of this additional transition operator.  First we should look back at the derivation of the exact matrix element  (\ref{reactamplvol1}).  In this  equation the potential describing $n-A$ scattering in the initial state is real and coincides with the potential supporting $(n\,A)$ bound state. Hence, these potentials cancel out each other.  However,  after we replace the exact three-body scattering wave function by the wave function $\Psi_{i}^{CDCC(+)}\,\varphi_{A}$  the initial $n-A$ potential becomes  complex while the final state $n-A$ potential is the real mean-field neutron  potential supporting the bound state. Replacing the  $n-A$ potential in the initial state by the energy-dependent one makes the problem of solving the CDCC equations difficult and impractical.  That is why  in practical applications  the adopted initial $n-A$ potential is complex local energy-independent one.  The conventional CDCC amplitude (\ref{CDCCpostconv1}) can be derived  from Eq. (\ref{reactamplvol1})  by using  the substitution $\Psi_{i}^{(+)} \to \Psi_{i}^{CDCC(+)}\,\varphi_{A}$, where the CDCC wave function satisfies Eq. (\ref{CDCC1}),   and  $\, V_{pA} +  V_{pn}  - U_{PF}\,$  by  $\,U_{pA}  + V_{pn} - U_{pF}$.   However,  a different expression for the CDCC amplitude  can be obtained if  we  start its derivation from equation
\begin{align}                         
{\mathcal M}^{(post)}  = <\Phi_{f}^{(-)} \big|V_{pA} + V_{pn} + V_{nA} - V_{nA} -   U_{pF}\big| \Psi_{i}^{(+)}> \Big|_{X \leq X_0},  
\label{volmatrel2}  
\end{align}
which is identical to  Eq. (\ref{reactamplvol1})   but in which we have not yet canceled out  $V_{nA}$  potentials.  The potential $(+V_{nA})$  in the transition operator  comes from the  Schr\"odinger equation for $\Psi_{i}^{(+)}$  and $(- V_{nA})$ from the Schr\"odinger equation for $\Phi_{f}^{(-}$.   If we  use the substitutions $\Psi_{i}^{(+)} \to \Psi_{i}^{CDCC(+)}\,\varphi_{A}$, $\,V_{pA} \to  U_{pA}$   and $(+ V_{nA}) \to (+ U_{nA})$  in Eq. (\ref{volmatrel2})  we get  Eq. (\ref{reactamplCDCCf2}), which can be transformed to the surface integral over ${\bm \rho}_{pF}$ rather than  the conventional one given by Eq. (\ref{CDCCpostconv1}).  

Now we can rewrite 
\begin{align}                         
{\mathcal M}_{conv}^{CDCC(post)}  \,=\, {\mathcal M}_{surf}^{CDCC(post)}   -  {\mathcal M}_{aux}^{CDCC(post)}
\label{CDCC2v1},
\end{align}
where  ${\mathcal M}_{conv}^{CDCC(post)}$ is the conventional CDCC stripping amplitude given by Eq. (\ref{CDCCpostconv1})
and 
\begin{widetext}
\begin{align}
&{\mathcal M}_{aux}^{CDCC(post)} =  \,<\chi_{pF}^{(-)}\,I_{A}^{F} \big|U_{nA}  -  V_{nA}^{sp} \big| \Psi_{i}^{CDCC(+)}>\Big|_{  \rho_{pF} \leq {\mathcal R}_{pF};\, r_{nA} \leq R_{nA} }
\label{CDCCaux1} \\
 &=i\,<\chi_{pF}^{(-)}\,I_{A}^{F} \big|{\rm Im}\,U_{nA} \big| \Psi_{i}^{CDCC(+)}>\Big|_{ \rho_{pF} \leq {\mathcal R}_{pF}; \,r \leq R_{nA}}.
\label{CDCCaux2}
\end{align}
\end{widetext} 
is the auxiliary amplitude.  Eq. (\ref{CDCCaux2})  follows from Eq. (\ref{CDCCaux1})   assuming that ${\rm Re} U_{nA}= V_{nA}^{sp}$.  Thus there is an ambiguity in the defintion of the CDCC amplitude.  If  we replace the exact scattering wave function by the CDCC one in the volume matrix element (\ref{volmatrel1})  we obtain the conventional CDCC reaction amplitude (\ref{CDCCpostconv1}). However,  if  we approximate  the exact scattering  wave function by the CDCC one in  the surface integral matrix element  (\ref{CDCCsurf1})  we obtain the  amplitude in the surface integral formalism ${\mathcal M}_{surf}^{CDCC(post)}$, which differs from  the conventional reaction amplitude ${\mathcal M}_{conv}^{CDCC(post)}$ by the auxiliary matrix element ${\mathcal M}_{aux}^{CDCC(post)}$, see Eq. (\ref{CDCC2v1}).

The ambiguity in the defintion of the CDCC amplitude is related with  the matrix element taken from the transition operator 
$U_{nA} - V_{nA}^{sp}$. The source of this ambuguity is the inconsistency in the treatment of the $n-A$  potentials when the many-body problem is reduced to the three-body one: to describe the $n-A$ interaction in the initial state the optical $U_{nA}$ is used while  the real potential $V_{nA}$ is adopted for describing the bound state $(nA)$  (see  Appendix  \ref{VnAUnADWBA1}, where  we discuss how the inconsistency in the treatment of the $n-A$ potential affects even the DWBA, which is more simpler than the CDCC).
This inconsistency remains an open question in the contemporary nuclear reaction theory if we use energy-independent $N-A$ potentials when reducing the many-body problem  to the three-body one.  
A similar problem appears in the treatment of the deuteron stripping reactions  using the Faddeev formalism in the momentum space, in which the integration over the energy requires energy-dependent nucleon-target optical potentials. These potentials  should provide scattering phase shifts at positive $N-A$ relative energies and possible bound states at negative relative energies.  

The replacement of the exact scattering wave function by the CDCC one is more accurate when it is done in the volume matrix element rather then in the surface one.  The volume matrix element is contributed by the internal and  peripheral (over the variable $r_{nA}$)  parts. While at low energy the external part dominates with energy increase  the role of the internal part also increases.  Meantime  the surface matrix element is mostly peripheral.  It is evident from the following consideration.
For large $\rho_{pF} \sim 30$ fm and small nonlocality $\,|{\mathcal R}_{pF} - {\mathcal R}_{dA}|\,$ of the post  form (see calculations  in section \ref{Calculations1})  $\,\rho_{dA}\,$ is also large.  Even if the initial CDCC wave function contains the $p-n$  pair in the continuum, the constraint over $l_{pn}$ constrains also the distance $r_{pn}$. Hence large $r_{nA}$ become dominant in the surface matrix element. Meantime the auxiliary matrix element is entirely contributed by the internal region because of the presence of $\,{\rm Im}U_{nA}$.  Thus the conventional amplitude is contributed by the internal auxiliary amplitude and mostly peripheral surface matrix element. 
Thus we suggest to use Eq. (\ref{CDCC2v1})  as the post CDCC amplitude, which can be expressed in terms of the predominantly peripheral surface matrix element and  the auxiliary amplitude.  

As we have underscored,  the constraint $X \leq X_0$ in the integration in the matrix elements in Eq. (\ref{CDCC2v1})  comes from the constraint in the exact matrix element (\ref{volmatrel2}).  The integrand in 
 ${\mathcal M}_{aux}^{CDCC(post)}$, which contains  the transition operator ${\rm Im}\,U_{nA}$, doesn't restrict the integration over $\rho_{pF}$, and the constraint $X \leq X_0$ comes only from the original matrix element  (\ref{volmatrel2}).  
That is why the amplitude ${\mathcal M}_{aux}^{CDCC(post)}$  may depend on the choice of $X_0$.  For peripheral reactions  the internal contribution in the post form is small  and  ${\mathcal M}_{aux}^{CDCC(post)}$ is also small  compared to ${\mathcal M}_{conv}^{CDCC(post)}$ because the depth of ${\rm Im}\,U_{nA}$  is significantly smaller than the depth of the real part of the transition operator in ${\mathcal M}_{conv}^{CDCC(post)} $ which is $\sim  V_{pn}$.  
Then the conventional CDCC amplitude  ${\mathcal M}_{conv}^{CDCC(post)}$  is close to the surface CDCC amplitude 
${\mathcal M}_{surf}^{CDCC(post)}$.  

Note that if we  use  the CDCC wave function satisfying the Schr\"odinger equation  \cite{moro2009}
\begin{align}
\big(E - T - U_{pA} - V_{nA}^{sp} -  V_{pn}\big)\,\Psi_{i}^{CDCC(+)}
=0, 
\label{CDCCmod1}
\end{align}
where the real $V_{nA}^{sp}$ is being used rather than the  optical potential  $U_{nA}$, then  
\begin{align}
{\mathcal M}_{surf}^{CDCC(post)} =  {\mathcal M}_{conv}^{CDCC(post)}= <\chi_{pF}^{(-)}\,I_{A}^{F} \big|U_{pA} +V_{pn} -  U_{pF}  \big| \Psi_{i}^{CDCC(+)}>\Big|_{X \leq X_0},
\label{CDCCconvmod1}
\end{align}
that is, the CDCC surface-integral form and the convential CDCC  amplitudes coincide.  However in this case the rearrangement channel $p  + (n\,A)$
is not suppressed and, hence solution of Eq.  (\ref{CDCCconvmod1}) is not unique.  For example, one can consider  $\Psi_{i}^{CDCC(+)} + \varphi_{nA}\,{\tilde \chi}_{pF}^{(+)}$, where  ${\tilde \chi}_{pF}^{(+)}$ is the $p-F$ distorted wave.  To decrease the contribution of the rearrangement channel the cut-off over $l_{pn}$  was introduced in \cite{moro2009}, however, the suppression of the rearrangement channels 
by the angular momentum cut-off is weaker than by the optical potentials \cite{austern88}. To achieve convergence the integration radius over $\rho_{pF}$ was extended up to 40 fm.  In \cite {moro2009} it  was also demonstrated that using of the CDCC wave function satisfying the Schr\"odinger equation  with the $U_{nA}$ optical potential rather than with $V_{nA}^{sp}$  gives the angular distribution  better agreeing  with the experimental one.    

We have expressed the conventional post CDCC amplitude ${\mathcal M}_{conv}^{CDCC(post)}\,$  given by the volume integral  in terms of the surface-integral matrix element ${\mathcal M}_{surf}^{CDCC(post)}$  and the internal auxiliary amplitude $\,{\mathcal M}_{aux}^{CDCC(post)}$.
There is no specific advantage of invoking the surface formalism   when we use the final channel wave function $\chi_{pF}^{(-)}\,I_{A}^{F}$ and the main goal here was to discuss the surface formalism just for better understanding of it.   However, below we will show another choice of the channel wave function, which clearly demonstrates the advantage of the surface formalism. 

\subsection{Prior form}
\label{priorformboundstae1}

Now we consider the prior form and derive the the reaction amplitude in the surface-integral formalism. 
We start from the exact prior form amplitude.

\subsubsection{Greider-Goldberger-Watson-Johnson choice of the final-channel wave function}

Here we consider a different choice of the exit channel wave function. We choose it to be a solution of the Schr\"odinger equation
\begin{align}
\big(E-K-  V_{pA} - V_{nA}\big)\,{\tilde \Phi}_{f}^{(-)} =0.
\label{GWJchwf1}
\end{align}
By comparing Eqs  (\ref{SchreqnPhif1}) and (\ref{GWJchwf1})  we can easily see the difference 
between the standard final channel wave function $\Phi_{f}^{(-)}$  and the newly defined  ${\tilde \Phi}_{f}^{(-)}$.      
Multiplying  Eq. (\ref{SchreqPsii1}) from the left by  ${\tilde \Phi}_{f}^{(-)*}$  and following a procedure similar to the one used for derivation of the exact reaction amplitude in the previous part we get
\begin{widetext}
\begin{align}                         
{\mathcal M}^{(post)}& = <{\tilde \Phi}_{f}^{(-)} \big|V_{pA} + V_{nA} + V_{pn} - V_{pA} - V_{nA}\big| \Psi_{i}^{(+)}> 
=<{\tilde \Phi}_{f}^{(-)} \big|V_{pn} \big| \Psi_{i}^{(+)}> \Big|_{r_{pn} \leq R_{pn}}     
\label{gwr1} \\
&=\,<{\tilde \Phi}_{f}^{(-)} \big|{\overleftarrow K} - {\overrightarrow K}\big| \Psi_{i}^{(+)}>\Big|_{r_{pn} \leq R_{pn}}.
\label{reactamplapp11}
\end{align}
\end{widetext}

The advantage of the new choice of the final channel wave function is that the transition operator is just $V_{np}$ and this keeps the nucleons of the deuteron within the range of their nuclear interaction. It allows  us to simplify the initial scattering wave function. However, the  new final channel wave function, a priori, cannot be factorized into a product of the $p-A$ distorted wave and the $n-A$ bound state wave function because now, owing to the presence of the $V_{pA}$, the recoil of the target can excite the system  $(nA)$ into any bound  or continuum  states.  As a result, the final channel wave function is contributed by the continuum component  $p+ n+ A$  and integration over $r_{dA}$  is not constrained.  The asymptotic behavior  of  $\,{\tilde \Phi}_{f}^{(-)*}\,$  at large $\,\rho_{pF}\,$
is given by the sum of the incident wave in the channel $p + F$ plus outgoing waves in all open two-body channels $p + F_{n}$, where $n$ denotes bound or excited states of $F$ plus three-body outgoing wave in the channel $p+n+A$. 
Converting the matrix element in Eq. (\ref{reactamplapp11})  containing ${\overleftarrow K} - {\overrightarrow K}$,
where $K= K_{pF} + K_{nA} + K_{A}$,  into surface integrals  we find that  only the integral over ${\rm {\bf \rho}}_{pF}$ survives giving the amplitude of the leading asymptotic term of the initial wave function in the rearrangement channel $p+F$.  Thus, using the surface integral formalism, it can be easily shown that the matrix element (\ref{gwr1})  coincides with the reaction amplitude for the stripping reaction $d+ A \to p + F$. The first proof of Eq. 
(\ref{gwr1})  was provided by Greider  \cite{greider}.  Although the final result was correct, the proof contained an error. 
The first correct proof of Eq. (\ref{gwr1})  was presented  by Goldberger and  Watson \cite{goldwatson}  and extensively used by Johnson and coworkers in the formulation of the ADWA  and its applications \cite{johnson70,johnsontandy74,johnson98,timofeyuk1999,pang2013}. 

For practical application we consider the limit  $A \to \infty$, in which the final channel wave function can be factorized  as 
\begin{align}
{\tilde \Phi}_{f}^{(-)} = \chi_{pA}^{(-)}\,\varphi_{F}.
\label{tildePhif1}
\end{align}
In this limit we can choose  ${\rm {\bf r}}_{pA}$ and ${\rm {\bf r}}_{nA}$  as two new  independent Jacobian variables.  Owing to the presence  of the bound state wave function $\varphi_{F}$ and  the potential $V_{np}$  as the transition operator,  the integration over both Jacobian variables is constrained.  The reaction amplitude is reduced to
\begin{align}                         
{\mathcal M}^{(post)} &= <\chi_{pF}^{(-)}\,\varphi_{F} \big|V_{pn} \big| \Psi_{i}^{(+)}> \Big|_{r_{nA} \leq {\mathcal R}_{nA};\,r_{pn} \leq R_{pn}}      
\label{mtrelVpn1}  \\                                                                                   
&=<\chi_{pF}^{(-)}\,\varphi_{F} \big|{\overleftarrow{K}} - {\overrightarrow{K}}\big| \Psi_{i}^{(+)}> \Big|_{r_{nA} \leq {\mathcal R}_{nA};\,r_{pn} \leq R_{pn}}                                                                      \nonumber\\
&=<\chi_{pF}^{(-)}\,\varphi_{F} \big|{\overleftarrow{K}}_{pA} - {\overrightarrow {K}}_{pA}   \big| \Psi_{i}^{(+)}> \Big|_{r_{nA} \leq {\mathcal R}_{nA};\,r_{pn} \leq R_{pn}}.
\label{gwr21} 
\end{align}
Here we took into account that $K_{nA}$ and $K_{A}$  are Hermitian operators because of the presence of the bound state wave function $\varphi_{F}$, that is integrating twice by parts we can transform ${\overleftarrow K}_{nA} +
{\overrightarrow K}_{A}$  to ${\overrightarrow K}_{nA} + {\overrightarrow K}_{A}$.  Because  ${\rm {\bf r}}_{pA}={\rm {\bf r}}_{nA} +{\rm {\bf r}}_{pn}$, limitation of the integration over ${\rm {\bf r}}_{pA}$ is
$r_{pA} \leq {\mathcal R}_{pA}= {\mathcal R}_{nA} + R_{pn}$. 

Now, as in the previous section, we approximate  the exact scattering wave function $\Psi_{i}^{(+)}$   by the CDCC one  $\Psi_{i}^{CDCC(+)}\,\varphi_{A}$ and replace the potential $V_{pA}$ in Eq. (\ref{GWJchwf1}) by the optical potential $U_{pA}$. As discussed previously, it can be done in the volume matrix element (\ref{mtrelVpn1}) containing  the transition operator $V_{pn}$ or in the matrix element containing $\,{\overleftarrow{K}}_{pA} - {\overrightarrow{K}}_{pA}$. The obtained amplitudes differ by the term containing the transition operator $U_{nA} - V_{nA}^{sp}$.  Actually,  if we do the approximation directly in the matrix element  (\ref{mtrelVpn1}) 
we get the conventional post CDCC amplitude  
\begin{align}                         
&{\mathcal M}_{conv}^{CDCC(post)}= 
<\chi_{pF}^{(-)}\,I_{F}\big|V_{pn} \big| \Psi_{i}^{CDCC(+)}\,> \Big|_{r_{nA} \leq {\mathcal R}_{nA};\,r_{pn} \leq R_{pn}}.    
\label{mtrelVpn3B1}                                                                           
\end{align}
Due to the presence of the short-range potential $V_{pn}$ we do not need to introduce an additional projector 
into the Schr\"odinger equation for the CDCC wave function, which constrains the distance between the proton and neutron 
(see Eq. (\ref{Ppnprojector1})  and Ref. \cite{austern87, moro2009}).   Another advantage of the presence of $\,V_{pn}\,$ is a possibility to 
approximate the CDCC wave function by the first term of the Weinberg states expansion \cite{pang2013}. The Weinberg states $\varphi_{i}^{W}$  are solutions of  the equation with eigenvalues $\lambda_{i}$:
\begin{align}
\big(-\varepsilon_{pn}^{d} - K_{pn} - \lambda_{i}\,V_{pn}\big)\,\varphi_{i}^{W}({\rm {\bf r}}_{pn})=0,
\qquad  i=1,2...
\label{Weinbergeq1}
\end{align}
This expansion significantly simplifies the calculation of the initial state scattering wave function.

Now we approximate $\Psi_{i}^{(+)}$ by the CDCC wave function $\Psi_{i}^{CDCC(+)}\,\varphi_{A}\,$ in the matrix element 
(\ref{gwr21})  to obtain the deuteron stripping amplitude in the surface-integral formalism:
\begin{widetext}
\begin{align}
&{\mathcal M}_{surf}^{CDCC(post)}= <\chi_{pF}^{(-)}\,I_{F}\big|{\overleftarrow K}_{pA} - {\overrightarrow K}_{pA}  \big| \Psi_{i}^{CDCC(+)}> \Big|_{r_{nA} \leq {\mathcal R}_{nA};\,r_{pn} \leq R_{pn}}
\label{surfKpA1}  \\
&=\,- \frac{{\mathcal R}_{pA}^{2}}{2\,\mu_{pA}} \int\,{\rm d}\,{\rm {\bf r}}_{nA}\,I_{A}^{F*}({\rm {\bf r}}_{nA})\,\int{\rm d}\,\Omega_{{\rm {\bf r}}_{pA}}\,
\Big[\Psi_{i}^{CDCC(+)}({\rm {\bf r}}_{dA},\,{\rm {\bf r}}_{pn})\,\frac{
\partial\,\chi_{pA}^{(-)*}({\rm{\bf r}}_{pA})\,}{\partial\,r_{pA}}                \nonumber\\
&- \chi_{pA}^{(-)*}({\rm {\bf  r}}_{pA})\,\frac{
\partial\,\Psi_{i}^{CDCC(+)}({\rm {\bf r}}_{dA},\,{\rm {\bf r}}_{pn})}{\partial\,r_{pA}} \Big]\Big|_{r_{pA}={\mathcal R}_{pA}; r_{pn} \leq  R_{pn}}.
\label{surfVpnCDCC1}
\end{align}
\end{widetext}
In this representation the matrix element  is actually the surface integral in the subspace over ${\rm {\bf r}}_{pA}$ and the volume integral over ${\rm {\bf r}}_{nA}$.  The main advantage of the surface-integral form  is that it is completely peripheral over $r_{pA}$  and  $r_{nA}$.  We take into account  that in the volume matrix element (\ref{gwr21})  the integration over $r_{pA}$ is limited by $\,r_{pA} \leq {\mathcal R}_{pA}$, where $\,{\mathcal R}_{pA}= {\mathcal R}_{nA} + R_{pn}$.   Actually we can take the surface  integral at any $r_{pA} > {\mathcal R}_{pA}$  but we do not want to do it because with the increasing of  the integration radius in the surface integral  we risk to be in the region where the CDCC wave function is not applicable.  So it is better to use the minimally required integration radius, which is ${\mathcal R}_{pA}$.
At $r_{pA}= {\mathcal R}_{pA}$  we make the integration over $r_{nA}$ peripheral.  From ${\rm {\bf r}}_{nA} = {\rm {\bf r}}_{pA} - {\rm {\bf r}}_{pn}$  follows  that  ${\mathcal R}_{pA} - R_{pn} \leq r_{nA}  \leq {\mathcal R}_{pA} + R_{pn}$.  Taking into account that  ${\mathcal R}_{pA} \sim 25-30$ fm  and that  $R_{pn}$ is small  we conclude that  $R_ {nA}  \leq r_{nA}  \leq {\mathcal R}_{nA} $,  where ${\mathcal R}_{nA} = {\mathcal R}_{pA}  - R_{pn}$ and $R_{nA}$ is the $n-A$ nuclear interaction radius. 
 
At  $r_{nA} \geq R_{nA}$ the radial overlap function can be replaced by its asymptotic term. We remind that the overlap function can be written as
\begin{align}
I_{A}^{F}({\rm {\bf r}}_{nA}) =& \sum\limits_{j_{nA}\,m_{j_{nA}\,m_{l_{nA}}}}\,<J_{A}\, M_{A}\,\, j_{nA}\, m_{j_{nA}}| J_{F}\,M_{F}> \nonumber\\
&\times <J_{n}\,M_{n}\,\,l_{nA}\,m_{l_{nA}}|j_{nA}\,m_{j_{nA}}>
 Y_{l_{nA}\,m_{l_{nA}}}({\rm {\bf {\hat r}}}_{nA})\,I^F_{A\,j_{nA}\,l_{nA}}(r_{nA}).
\label{overlapfunction2}
\end{align}    
Here, $l_{nA}\,(m_{l_{nA}})$  is the relative orbital angular momentum (its projection) of  $n-A$  in the bound state
$F=(n\,A)$,  $j_{nA}\, (m_{j_{nA}})$ is the total angular momentum (its projection) of the neutron in the bound state,  $J_{i}\,(M_{i})$  is the spin (its projection) of nucleus $i$.  The radial overlap function at $r_{nA} > R_{nA}$ 
takes the form 
\begin{widetext}
\begin{align}
&I_{A\,\,j_{nA}\,l_{nA}}^{F}(r_{nA}) \stackrel{r_{nA}\leq R_{nA}}{\approx} C^{F}_{A\,\,j_{nA}\, l_{nA}}\,i^{l_{nA}+1} \,\kappa_{nA}\,h_{l_{nA}}^{(1)}(i\,\kappa_{nA}\,r_{nA})        
 \stackrel{r_{nA} \to \infty}{\approx} C^{F}_{A\,\,j_{nA}\, l_{nA}}\,\, \frac{e^{-\kappa_{nA}\,r_{nA}}}{r_{nA}},
\label{radovfunctasympt1} 
\end{align}
where $h_{l_{nA}}^{(1)}(i\,\kappa_{nA}\,r_{nA})$ is the spherical Hankel function of the first order, 
$C^{F}_{A\,\,j_{nA}\, l_{nA}}$ is the ANC of the overlap function, 
$\kappa_{nA}= \sqrt{2\,\mu_{nA}\,\varepsilon_{nA}^{F}}\,$  is the bound state wave number  and $\varepsilon_{nA}^{F}$ 
is the binding energy of the ground state of $F$ for the virtual decay $ F \to n+ A$.
Taking into account Eqs (\ref{overlapfunction2})  and  (\ref{radovfunctasympt1})   we get the final expression 
for the post-form CDCC deuteron stripping amplitude in the surface-integral formalism:
\begin{align}
{\mathcal M}_{surf}^{CDCC(post)}= & - \sum\limits_{j_{nA}\,m_{j_{nA}\,m_{l_{nA}}}}\,<J_{A}\, M_{A}\,\, j_{nA}\, m_{j_{nA}}| J_{F}\,M_{F}> \,<J_{n}\,M_{n}\,\,l_{nA}\,m_{l_{nA}}|j_{nA}\,m_{j_{nA}}>  \nonumber\\
&\times i^{-l_{nA}-1} C^{F }_{A\,\,j_{nA}\, l_{nA}}\,\,\kappa_{nA}\,\frac{{\mathcal R}_{pA}^{2}}{2\,\mu_{pA}} \int\limits_{ R_{nA} \leq  r_{nA} \leq {\mathcal R}_{nA}}\,{\rm d}\,{\rm {\bf r}}_{nA}\,Y_{l_{nA}\,m_{l_{nA}}}^{*}({\rm {\bf {\hat r}}}_{nA})\,h_{l_{nA}}^{(1)*}(i\,\kappa_{nA}\,r_{nA})
\nonumber\\
&\times
\int{\rm d}\Omega_{{\rm {\bf r}}_{pA}}\,\Big[\Psi_{i}^{CDCC(+)}({\rm {\bf r}}_{dA},\,{\rm {\bf r}}_{pn})\,\frac{
\partial\,\chi_{pA}^{(-)*}({\rm{\bf r}}_{pA})\,}{\partial\,r_{pA}}                \nonumber\\
&- \chi_{pA}^{(-)*}({\rm {\bf  r}}_{pA})\,\frac{
\partial\,\Psi_{i}^{CDCC(+)}({\rm {\bf r}}_{dA},\,{\rm {\bf r}}_{pn})}{\partial\,r_{pA}} \Big]\Big|_{r_{pA}={\mathcal R}_{pA}; r_{pn} \leq  R_{pn}}.
\label{surfVpnCDCCfinal1}
\end{align}
\end{widetext}

Thus the original volume matrix element can be converted into the surface integral over $\,r_{pA}\,$, which, due to the constraint on the variable $r_{pn}$,  leads to the dominant contributions for $\,r_{nA} \geq R_{nA}$. It allows us to parameterize the reaction amplitude in terms of the ANC. This peripheral character of the reaction amplitude is obtained because we used  the modified final channel wave function.

We can relate now the ${\mathcal M}_{surf}^{CDCC(post)}$  and the conventional CDCC amplitude  ${\mathcal M}_{conv}^{CDCC(post)}$.  To this end  we rewrite (\ref{mtrelVpn3B1})  as
 \begin{widetext}
\begin{align}                         
{\mathcal M}_{conv}^{CDCC(post)} =& 
<\chi_{pF}^{(-)}\,I_{F}\big|U_{pA} + U_{nA} + V_{pn}- U_{pA} - V_{nA}^{sp} 
\nonumber \\ &
+  [V_{nA}^{sp} - U_{nA}]  \big| \Psi_{i}^{CDCC(+)}> \Big|_{r_{nA} \leq {\mathcal R}_{nA};\,r_{pn} \leq R_{pn}}    
                    \nonumber\\
=&  {\mathcal M}_{surf}^{CDCC(post)} - {\mathcal M}_{aux}^{CDCC(post)} .
\label{mtrelCDCCsurfVpn1}                                                                           
\end{align}
Thus, as before we can rewrite the conventional post CDCC volume matrix element in terms of two amplitudes: the entirely peripheral surface-integral matrix element and the internal auxiliary one. 
The matrix element in the surface-integral form is expressed in terms of the potential transition operator
\begin{align}
{\mathcal M}_{surf}^{CDCC(post)}&= <\chi_{pF}^{(-)}\,I_{F}\big|U_{pA} + U_{nA} + V_{pn}- U_{pA} - V_{nA}^{sp} \big| \Psi_{i}^{CDCC(+)}> \Big|_{r_{nA} \leq {\mathcal R}_{nA};\,r_{pn} \leq R_{pn}}  
\label{surfpot1}  \\
&=<\chi_{pF}^{(-)}\,I_{F}\big|{\overleftarrow K} - {\overrightarrow K}  \big| \Psi_{i}^{CDCC(+)}> \Big|_{r_{nA} \leq {\mathcal R}_{nA};\,r_{pn} \leq R_{pn}}.
\label{surfK11} 
\end{align}
When deriving (\ref{surfK11})  we took into account that $\Psi_{i}^{CDCC(+)}$  satisfies the Schr\"odinger equation with the potential $U_{pA} + U_{nA} + V_{pn}$ and the final channel wave function is the solution of the Schr\"odinger equation with  the potential $U_{pA} + V_{nA}^{sp}$.  It allows  us to replace    
 $U_{pA} + U_{nA} + V_{pn} - U_{pA} - V_{nA}^{sp}$  in the matrix element  (\ref{surfpot1}) by ${\overleftarrow K} - {\overrightarrow K}$  what leads to the surface matrix element  (\ref{surfVpnCDCCfinal1}). The auxiliary matrix element, which is entirely contributed by the nuclear interior, is written as
\begin{align}                         
&{\mathcal M}_{aux}^{CDCC(post)}= 
<\chi_{pF}^{(-)}\,I_{F}\big| [ U_{nA} -  V_{nA}^{sp}]  \big|  \Psi_{i}^{CDCC(+)}> \Big|_{r_{nA} \leq R_{nA};\,r_{pn} \leq R_{pn}}      \nonumber\\
&=<\chi_{pF}^{(-)}\,I_{F}\big| {\rm Im}\,U_{nA} \big|  \Psi_{i}^{CDCC(+)}> \Big|_{r_{nA} \leq R_{nA};\,r_{pn} \leq R_{pn}}   
\label{mtrelCDCCsVpnaux1}   .                                                                        
\end{align}
\end{widetext}
In Eq. (\ref{mtrelCDCCsVpnaux1}) we adopted  ${\rm Re}\,U_{nA}= V_{nA}^{sp}$. 
We remind that  the auxiliary matrix element  ${\mathcal M}_{aux}^{CDCC(post)}$  appears due to the inconsistency in treating  the  $n-A$ potential. The auxiliary matrix element is contributed by the range of the imaginary part of $U_{nA}$  potential,  that is  $r_{nA}\leq R_{nA}$.  The depth of the imaginary part  of  $\,U_{nA}\,$ is significantly smaller than that of $\,V_{pn}$. Also, the constraint $r_{pn} \leq R_{pn}$ keeps protons in the region with the strongest absorption.  Hence  we expect that  $|{\mathcal M}_{aux}^{CDCC(post)}|$  can be significantly smaller  than $\,|{\mathcal M}_{conv}^{CDCC(post)}|$  at low energies and good matching of the initial and final momenta. In this case 
\begin{align}
{\mathcal M}_{conv}^{CDCC(post)}  \approx  {\mathcal M}_{surf}^{CDCC(post)}.
\label{convsurfCDCC1}
\end{align}

Once again we repeat that adoption of the Greider-Goldberger-Watson-Johnson final channel wave function allowed us to constrain the integration over $\,r_{pn}\,$ by the range of the transition operator $\,V_{pn}\,$ despite the fact that the CDCC wave function contains the  components describing the $\,p-n\,$ pair in the continuum.  As we mentioned, it allows one  to  approximate the CDCC wave function by the first term of the Weinberg states expansion \cite{pang2013} and this significantly simplifies the calculation of the initial state scattering wave function.  

The presence of the overlap function $\,I_{A}^{F}\,$ constrains the integration over $r_{nA}$. As the result of these two constraints the surface matrix element  taken at $\,r_{pA}= {\mathcal R}_{pA}\,$ leads to the dominant contribution at $\,r_{nA} \geq R_{nA}$. In other words, the surface matrix element is peripheral allowing us to parametrize it in terms of the ANC for the bound state $F=(n\,A)$, which is the only  model-independent spectroscopic information extractable from experiment \cite{muk2009}.  The auxiliary term determines the contribution from the nuclear interior.
Although here equations were obtained assuming an infinitely heavy target $A$, they should work also for a heavy target with a finite mass. Necessary  corrections may be introduced using expansion over a small parameter $1/A$.

\section{Deuteron stripping to a resonance state}
\label{resonancestates1}

Now we proceed to the  main goal of the present paper and apply the surface formalism used in the previous sections for stripping to bound states, to describe the deuteron stripping populating resonance states. 

\subsection{Prior form}
\label{priorformresonancestate1}

To treat the stripping to resonance states we use the prior formalism, in which the exact scattering wave function $\,\Psi_{f}^{(-)}\,$ is taken in the final
state. We consider the deuteron stripping reaction
\begin{align}
d + A  \to p + n + A,
\label{deutstr1}
\end{align}
proceeding through  the  resonant sub-reaction $n + A \to F^{*} \to n + A$.  
The results  can easily be extended for the deuteron stripping reaction
\begin{align}
d + A \to  p + b + B,
\label{deutrstr2}
\end{align}
which proceeds through the resonant sub-reaction $n + A \to F^{*} \to b + B$, where the channel $b + B$ differs from $n + A$. 

The wave function $\,\Psi_{f}^{(-)}\,$ satisfies the Schr\"odinger equation
\begin{align}
\Psi_{f}^{(-)*}\,\big(E - {\overleftarrow K} - V_{pA} - V_{nA} - V_{pn} - H_{A} \big)\,=\,0
\label{shreqPsif1}
\end{align}
and has the  $\,p + n + A\,$ incident  three-body  wave in the continuum with the outgoing waves in all the open channels. Let $\,\Phi_{i}^{(+)} = \varphi_{pn}\,\chi_{dA}^{(+)}\,$ be the wave function of the entry channel, $\,\chi_{dA}^{(+)}\,$ be the $\,d +A\,$  distorted wave.  
We adopt  the initial channel wave function as the solution of the Schr\"odinger equation
\begin{align}
\big(E- K - V_{pn} - U_{dA} - H_{A} \big)\,\Phi_{i}^{(+)}\,=\,0
\label{shreqPhii1}
\end{align}
with the $d+ A$ incident wave.

Multiplying  Eq. (\ref{shreqPsif1}) from the right  by $\,\Phi_{i}^{(+)}\,$  and taking into account  Eq. (\ref{shreqPhii1})   we get
\begin{align}
&{\mathcal M}^{(as)} \,= \,<\Psi_{f}^{(-)}\big| {\overleftarrow  K} -  {\overrightarrow K}\big| \Phi_{i}^{(+)}>
\label{exprsur1}    \\
&=<\Psi_{f}^{(-)}\big| V_{pA} + V_{nA} + V_{pn} -  V_{pn} - U_{dA}\big| \Phi_{i}^{(+)}>        \nonumber\\
&= <\Psi_{f}^{(-)}\big| V_{pA} + V_{nA}  -  U_{dA}\big| \Phi_{i}^{(+)}> \, = \, {\mathcal M}^{(prior)}.
\label{exprvol1}
\end{align}
Eq. (\ref{exprvol1}) is the standard prior form of the volume matrix element, while Eq. (\ref{exprsur1}) is the matrix element, which can be written  in a surface-integral form. This matrix element can be easily reduced  to the amplitude of the leading asymptotic term  of the $\,\Psi_{f}^{(-)*}\,$ in the channel $\,d+ A$.  This amplitude, by definition, is the deuteron stripping amplitude  $\,{\mathcal M}^{(as)}$.  To show it  we rewrite
\begin{widetext}
\begin{align}
<\Psi_{f}^{(-)}\big| {\overleftarrow  K} -  {\overrightarrow K}\big| \Phi_{i}^{(+)}>\,= \, <\Psi_{f}^{(-)}\big| {\overleftarrow  K}_{dA} -  {\overrightarrow K}_{dA}\big| \Phi_{i}^{(+)}>
+ <\Psi_{f}^{(-)}\big| {\overleftarrow  K}_{pn} -  {\overrightarrow K}_{pn}\big| \Phi_{i}^{(+)}>.
\label{surfmatrel1}
\end{align}

The matrix element containing ${\overleftarrow K}_{pn} - {\overrightarrow  K}_{pn}$ 
vanishes because it contains the deuteron bound state wave function $\varphi_{pn}$.
Taking the limit $R_{pn} \to \infty$ we get
\begin{align}
<\Psi_{f}^{(-)} \Big|{\overleftarrow K}_{pn} - {\overrightarrow  K}_{pn}\Big|\,\Phi_{i}^{(+)}>=&  -  \lim \limits_{R_{pn} \to \infty} \frac{R_{pn}^{2}}{2\,\mu_{pn}}\,\int\,{\rm d}{\rm {\bm \rho}}_{dA}\,\chi_{dA}^{(+)}({\rm {\bm \rho}}_{dA})\,\int\,{\rm d}\Omega_{ {\rm {\bf { r}}}_{pn}}
\nonumber \\ & \times
\Big[\Psi_{f}^{(-)*}({\rm {\bm \rho}}_{dA},{\rm {\bf r}}_{pn})\frac{{\partial \varphi_{pn}(r_{pn})}}{{\partial {r_{pn}}}}                            
- \varphi_{pn}(r_{pn})\frac{{\partial \Psi_{f}^{(-)*}({\rm {\bm \rho}}_{dA},{\rm {\bf r}}_{pn})}}{{\partial {r_{pn}}}}\Big]\Big|_{r_{pn}= R_{pn}}   \nonumber \\  = & 0.
\label{matrelemrpn1}
\end{align}
Hence, 
\begin{align}
&{\mathcal M}^{(as)}\, = \,<\Psi_{f}^{(-)}\big| {\overleftarrow  K} -  {\overrightarrow K}\big| \Phi_{i}^{(+)}>= <\Psi_{f}^{(-)}\big| {\overleftarrow  K}_{dA} -  {\overrightarrow K}_{dA}\big| \Phi_{i}^{(+)}>                                   \nonumber\\
=&  - \,  \lim \limits_{R_{dA} \to \infty}\,\frac{R_{dA}^{2}}{2\,\mu_{dA}}\,\int\,{\rm d}{\rm {\bf  r}}_{pn}\,\varphi_{pn}({\rm {\bf r}}_{pn})\,\int\,{\rm d}\Omega_{ {\rm {\bm { \rho}}}_{dA}}
\nonumber \\ & \times
\Big[\Psi_{f}^{(-)*}({\rm {\bm \rho}}_{dA},{\rm {\bf r}}_{pn})\frac{{\partial \chi_{dA}^{(+)}({\rm {\bm \rho}}_{dA})}}{{\partial {\rho_{dA}}}}                            
- \chi_{dA}^{(+)}({\rm {\bm \rho}}_{dA})\frac{{\partial \Psi_{f}^{(-)*}({\rm {\bm \rho}}_{dA},{\rm {\bf r}}_{pn})}}{{\partial {\rho_{dA
}}}}\Big]\Big|_{\rho_{dA}= R_{dA}}                            \nonumber\\
=&  {\mathcal  M}^{prior}.
\label{surfmatreldA1}
\end{align}
 \end{widetext} 
To prove that  this equation  reduces to $\,{\mathcal M}^{(as)}\,$  we have taken into account that  at $\,\rho_{dA}  \to \infty\,$ only the leading asymptotic term 
\begin{align}
\Psi_{f}^{(-)*}({\rm {\bm \rho}}_{dA},\,{\rm {\bf r}}_{pn})  \sim  - \frac{\mu_{dA}}{2\,\pi}\,{\mathcal M}^{(as)}\,u^{+}( \rho_{dA})\,\varphi_{pn}\,\varphi_{A},
\label{asPsifdA1}
\end{align}
where $\,u^{+}(\rho_{dA})\,$  is outgoing scattered wave in the $\,d+A\,$ two-body channel,  will give non-vanishing contribution to the  integral over  $\bm \rho_{dA}$.  Thus in the prior form the conventional CDCC amplitude given by the volume matrix element is equal to the amplitude in the surface-integral formalism. It is because only one potential, $V_{nA}^{sp}$, is used in the prior formalism.   

After proving that the volume matrix element (\ref{exprvol1}) is equal to the amplitude of the total scattering wave function in the asymptotic final $\,d + A \,$ channel ${\mathcal M}^{(as)}$, we consider now the constraints on the integration volume in the matrix element  (\ref{exprvol1}). 
Owing to the presence of the deuteron bound state wave function in the initial channel, the integration over $\,r_{pn}\,$ is limited.
At large $\,r_{pA}\,$  $\,V_{pA} \to U_{pA}^{C}\,$, where $\,U_{pA}^{C}= Z_{A}\,e^{2}/r_{pA}\,$ is the  Coulomb potential between the proton and the center of mass of nucleus $\,A$;   also at large $\,r_{pA}\,\,$ $\,U_{dA} \to U_{dA}^{C}\,$ because at large $\,r_{pA}\,$ also $\,r_{dA}\,$ is large because of the constrain of $\,r_{pn}\,$. For the same reason when $\,r_{pA}\,$ increases also $\,r_{nA}\,$ increases. Then $\,V_{nA}\,$ vanishes when $\,r_{pA}\,$ increases. As $r_{dA}$ increases the matrix element from the difference  $U_{pA}^{C} - U_{dA}^{C}$ goes to zero as $d_{0}\,Z_{A}\,e^{2}/(2\,r_{dA}^{2})$, where $d_{0}$ is the deuteron size \cite{bunakov1970}. Hence the integration over $r_{dA}$ is also constrained. Thus the volume integral in Eq. (\ref{exprvol1}) 
can be taken over the constrained volume in the 6-dimensional space $\big\{{ {\bm \rho}}_{dA},\,{\rm {\bf r}}_{pn}\big\}$ with the hyper-radius   $Y \le Y_0$, where 
$Y_0= ({\mu_{pn}}\, {\mathcal R}_{pn}^{2}/m + { \mu_{dA}}\, {\mathcal R}_{dA}^{2}/m)^{1/2}$.
Also
 ${\mathcal R}_{pn} $ is  the maximal $r_{pn}$, which is required to achieve a desired accuracy for the integral over $r_{pn}$ and ${\mathcal R}_{dA} $ is  the maximal $\rho_{dA}$, which is required to achieve a desired accuracy for the integral over $\rho_{dA}$.

Hence we can rewrite (\ref{exprvol1})  in form of the conventional  volume  and the surface-integral forms:
\begin{align}
&{\mathcal M}^{(prior)} \,= <\Psi_{f}^{(-)}\big| V_{pA} + V_{nA}  -  U_{dA}\big| \Phi_{i}^{(+)}>\Big|_{ \rho_{dA} \leq {\mathcal R}_{dA};\,   r_{pn} \leq {\mathcal R}_{pn} }    
\label{exprvolconstr1}                                                       \\
&=  -  \frac{R_{dA}^{2}}{2\,\mu_{dA}}\,\int\,{\rm d}{\rm {\bf  r}}_{pn}\,\varphi_{pn}({\rm {\bf r}}_{pn})\,\int\,{\rm d}\Omega_{ {\rm {\bm { \rho}}}_{dA}}
\nonumber \\ 
& \times  \Big[\Psi_{f}^{(-)*}({\rm {\bm \rho}}_{dA},{\rm {\bf r}}_{pn})\frac{{\partial \chi_{dA}^{(+)}({\rm {\bm \rho}}_{dA})}}{{\partial {\rho_{dA}}}}                            
- \chi_{dA}^{(+)}({\rm {\bm \rho}}_{dA})\frac{{\partial \Psi_{f}^{(-)*}({\rm {\bm \rho}}_{dA},{\rm {\bf r}}_{pn})}}{{\partial {\rho_{dA
}}}}\Big]\Big|_{ \rho_{dA} = {\mathcal R}_{dA};\,   r_{pn} \leq {\mathcal R}_{pn} }\\
={\mathcal  M}^{(as)} . 
\label{surfintexconstr1}
\end{align}
Thus the integration in both forms, volume and surface, is constrained. 

As in the previous sections, now  we can get the prior form of the CDCC amplitude  for the stripping to the resonance state in the conventional volume integral form and the surface formalism. To this end we replace $\Psi_{f}^{(-)}$  by the CDCC wave function $\Psi_{f}^{CDCC(-)}\,\varphi_{A}$.  If we do it in the matrix element  (\ref{exprvolconstr1}) containing the volume integral we get the conventional 
CDCC prior form amplitude:
\begin{align}
&{\mathcal M}_{conv}^{CDCC(prior)} \,= <\Psi_{f}^{CDCC(-)}\big| U_{pA} + V_{nA}^{sp}  -  U_{dA}\big| \varphi_{pn}\,\chi_{dA}^{(+)}>\Big|_{ \rho_{dA} \leq {\mathcal R}_{dA};\,   r_{pn} \leq {\mathcal R}_{pn} }.    
\label{CDCCvolconstr1}                                                       
\end{align}
To obtain the prior form of the CDCC matrix element we replaced  $<\varphi_{A}|V_{PA}|\varphi_{A}>$ by the optical potential $U_{pA}$.
Correspondingly, from  (\ref{surfintexconstr1})  we get the CDCC prior form amplitude  in the surface integral representation:
\begin{align}
&{\mathcal M}_{surf}^{CDCC(prior)} \,=\,  - \frac{R_{dA}^{2}}{2\,\mu_{dA}}\,\int\,{\rm d}{\rm {\bf  r}}_{pn}\,\varphi_{pn}({\rm {\bf r}}_{pn})\,\int\,{\rm d}\Omega_{ {\rm {\bm { \rho}}}_{dA}}
\nonumber \\ 
& \times  \Big[\Psi_{f}^{CDCC(-)*}({\rm {\bm \rho}}_{dA},{\rm {\bf r}}_{pn})\frac{{\partial \chi_{dA}^{(+)}({\rm {\bm \rho}}_{dA})}}{{\partial {\rho_{dA}}}}                            
- \chi_{dA}^{(+)}({\rm {\bm \rho}}_{dA})\frac{{\partial \Psi_{f}^{CDCC(-)*}({\rm {\bm \rho}}_{dA},{\rm {\bf r}}_{pn})}}{{\partial {\rho_{dA
}}}}\Big]\Big|_{ \rho_{dA} ={\mathcal R}_{dA};\, r_{pn} \leq {\mathcal R}_{pn} }. 
\label{surfCDCCconstr1}
\end{align}

Because the potential  $V_{nA}^{sp}$  is real  both conventional and surface integral forms  coincide.  It is straightforward to see  but before showing it we discuss the CDCC wave  function $\,\Psi_{f}^{CDCC(-)}\,$. We consider the deuteron stripping reaction populating a resonance state, which decays into the channel $n + A$.  Thus we have the three-body system  $p+n+A$ in the final state, in which we need to take into account explicitly the $n+ A$  rescattering  in the final state to describe the resonance in the $n-A$ system.  To this end in the finite volume around the target $A$   we approximate the exact final state scattering wave function by  the CDCC  wave  function 
$\Psi_{f}^{CDCC(-)}\,\varphi_{A}$, which satisfies the three-body Schr\"odinger equation
\begin{align}
\Psi_{f}^{CDCC(-)*}\,\big(E -   {\overleftarrow  K}  - U_{pA} - V_{nA}^{sp} - V_{pn}\big)=0.
\label{ShreqCDCCf1}
\end{align}
The CDCC method simplifies the problem by considering only one equation (\ref{ShreqCDCCf1})  with the incident wave describing the three-body system $p+ n + A$  in the continuum. The simplest mechanism of the  deuteron stripping populating a resonance state can be described as virtual breakup of the deuteron with subsequent $n + A$ resonance scattering, in which the proton is a spectator. 
An effective way to describe the three-body system in the continuum, which takes into account the resonance scattering in the sub-system $n+ A$,  is to use the CDCC wave function which is expressed in terms of the product of the $n-A$ scattering wave function times the scattering wave function of the proton off the center of mass of the system $n + A$.  Then the only channel  coupled to the three-body continuum that can be included in the CDCC method, is the two-fragment channel $p +F$, where $F=(n\,A)$ is the bound state. A few bound states of the system $\,(n\,A)\,$ can be taken into account. Then we can write the CDCC wave function in the form
\begin{widetext}
\begin{align}
\Psi_{f}^{CDCC(-)}({\rm {\bm \rho}}_{pF},\,{\rm {\bf r}}_{nA}) = \,\sum\limits_{i=0}^{i_{max}}\,\varphi_{nA}^{(i)}({\rm {\bf r}}_{nA})\,\chi_{{\rm {\bf q}}_{pF}}^{(i)(-)}({\rm {\bm \rho}}_{pF})\, + \,\sum\limits_{j=1}^{j_{max}}\, 
{\overline \psi}_{ {\rm {\bf  k}}_{nA} }^{(j)(-)}({\rm {\bf r}}_{nA})\,{\overline \chi}_{ {\rm {\bf q}}_{pF}({\bf k}_{nA})  }^{(j)(-)}({\rm {\bm \rho}}_{pF}),
\label{cdccwf1}
\end{align}
\end{widetext}
Here $\varphi_{nA}^{(i)}({\rm {\bf r}}_{nA})$ is the $i$-th  bound state wave function of the system $F=(nA)$ with $i=0$ corresponding to the ground state and $\chi_{{\rm {\bf q}}_{pF}}^{(i)(-)}({\rm {\bm \rho}}_{pF})$ are the functions, which describe the relative motion of the center-of-mass of $p$ and the $\,(n\,A)\,$ pair in the $i$-th bound state.  ${\overline \psi}_{{\rm {\bf  k}}_{nA} }^{(j)(-)}({\rm {\bf r}}_{nA})$ is the $n-A$ scattering wave function  obtained by averaging continuous breakup states in the $j$-th bin  and $\,{\overline \chi}_{{\rm {\bf q}}_{pF}({\bf k}_{nA}) }^{(j)(-)}({\rm {\bm \rho}}_{pF})\,$  is the wave function 
describing the relative motion of the proton and the center of mass of the system $n+ A$ in the continuum in the $j$-th bin. In Eq. (\ref{cdccwf1})   the relative momentum   $\,{\rm {\bf q}}_{pF}({\rm{\bf k}}_{nA})\,$ of  the  particles $p$ and $F$ is related to the $n-A$ relative momentum ${\rm {\bf k}}_{nA}$ via the energy conservation law:
\begin{align}
E= E_{dA} - \varepsilon_{pn}^{d} = E_{pF} - \varepsilon_{nA}^{F}= \frac{q_{pF}^{2}}{2\,\mu_{pF}} + \frac{k_{nA}^{2}}{2\,\mu_{nA}}.
\label{qpFknArel1}
\end{align}
The $n-A$ interaction is taken as a real single-particle  potential
$V_{nA}^{sp}= <\varphi_{A}|V_{nA}|\varphi_{A}>$ , which can support the resonance in the $n-A$ system. The corresponding scattering wave function is orthogonal to the bound states generated by this potential.  

In order to be sure that Eq. (\ref{cdccwf1}) provides a unique solution of Eq. (\ref{ShreqCDCCf1})  we need to suppress the two-fragment rearrangement channles,
$\,n + (pA)\,$ and $\,d + A$.   Unfortunately there is only one optical potential, $U_{pA}$,  in Eq. (\ref{ShreqCDCCf1}).    This potential  to some extent suppresses the rearrangement channel $ n + (p\,A)$ because  it  generates a substantial positive imaginary part  to the potential $n + (p\,A)$ damping the outgoing neutron wave.   However,  the other rearrangement channel $d+ A$ is not suppressed because the potential $\,V_{pn}\,$ is real.  
To  provide a unique solution  of  Eq. (\ref{ShreqCDCCf1})  a model space is introduced in which the CDCC solution becomes unique. This model space is achieved by cutting the  $n-A$ relative orbital angular momenta by some finite $l_{nA}^{max}$ \cite{sawada86}. Although the solution is unique in such a model space because the rearrangement channels are absent  in the asymptotic regions, the non-uniqueness  is disguised in the dependence of the CDCC solution on the adopted  model space  \cite{sawada86}.  Fortunately, in the case of the stripping to resonance the number  of the resonant  partial waves $\,l_{nA}\,$ is limited by one or a few at most. To ensure the uniqueness of the CDCC solution only the number of the non-resonant partial waves (non-resonant background) in the subsystem $n-A$ requires a cut-off  that can create a model dependence on $l_{nA}^{max}$. 
Note that  a constraint on $l_{nA}$ keeps $n$ close to $A$  suppressing  the contribution of the rearrangement channel $d + A$.   

We write down now the $n-A$ scattering wave function taking into  account  the spins in the representation with given channel spin and its projections: 
\begin{align}
& \psi _{{\rm {\bf k}}_{nA}s\,m_{s}\,m''_{s}}^{(j)( - )}({\rm {\bf r}}_{nA}) = \,i\,\frac{2\,\pi}{k_{nA}\,r_{nA}}\,\varphi_{A}\,
\sum\limits_{J_{F}\,M_{F}\,l_{nA}\,m_{l_{nA}}\,m''_{l_{nA}}}  \,<s\,{m_s}\,\,l_{nA}\,m_{l_{nA}}|J_{F}\,M_{F}>\,   \nonumber\\
&\times < s\, m''_{s} \,\,l_{nA}\,m''_{l_{nA}}|J_{F}\,M_{F}>\, i^{-l_{nA}}\,Y_{l_{nA}\,m_{l_{nA}}}^{*}({\rm {\bf {\hat k}}}_{nA})\,Y_{l_{nA}\,m''_{l_{nA}}}({\rm {\bf {\hat r}}}_{nA})                                                 
 \,\phi _{nA\,s\,m''_{s}}\,u_{k_{nA}\,s\,l_{nA}\,J_{F}}^{(j)(+)*}(r_{nA}).
\label{psielc11}
\end{align}
Here $s$ is the channel spin ($m_{s}$ and $m''_{s}$ are its projections before and after scattering) and $l_{nA}$ is the $n-A$ orbital angular momentum ($m_{l_{nA}}$ and $m''_{l_{nA}}$ are its projections before and after scattering), $\,J_{F}$ ($\,M_{F}$) is the spin (its projection) of nucleus $F$, $\phi_{nA\,s\,m''_{s}}$ is the spin function of the system $n +A$ with the channel spin $s$.  We presented here only the diagonal components (over the channel spin and the orbital angular momenta) of  the scattering wave function.    
General cases of the scattering wave function with different channel spins in the initial and final states  and even including  reaction channels are given in \cite{muk2011}.  

Note that in practical application we need to use ${\overline \psi} _{{\rm {\bf k}}_{nA}s\,m_{s}\,m''_{s}}^{(j)( - )*}({\rm {\bf r}}_{nA})$,
which is expressed in terms of the binned radial wave function  ${\overline u}_{k_{nA}\,s\,l_{nA}\,J_{F}}^{(j)(+)}(r_{nA})$ given by   \cite{thompson}
\begin{align}
{\overline u}_{k_{nA}\,s\,l_{nA}\,J_{F}}^{(j)(+)}(r_{nA})  = \sqrt{\frac{2}{\pi\,N_{s\,l_{nA}\,J_{F}}^{(j)}}}\,\int\limits_{k_{nA}^{(j-1)}}^{k_{nA}^{(j)}}\,{\rm d}k_{nA}\,g_{s\,l_{nA}\,J_{F}}^{(j)}(k_{nA})\,u_{k_{nA}\,s\,l_{nA}\,J_{F}}^{(+)}(r_{nA}) ,
\label{binnA1}
\end{align}
where  $g_{s\,l_{nA}\,J_{F}}^{(j)}(k_{nA})$  is the weight function. The normalization constant is
\begin{align}
N_{s\,l_{nA}\,J_{F}}^{(j)}= \int\limits_{k_{nA}^{(j-1)}}^{k_{nA}^{(j)}}\,{\rm d}\,k_{nA}\, |g_{s\,l_{nA}\,J_{F}}^{(j)}(k_{nA})|^{2}.
\label{normalizconst1}
\end{align}
The adopted normalization constant  $\,N_{s\,l_{nA}\,J_{F}}^{(j)}$  makes an orthonormal set ${\overline u}_{k_{nA}\,s\,l_{nA}\,J_{F}}^{(j)(+)*}(r_{nA}) $
when all the intervals $(k_{nA}^{j-1},\,k_{nA}^{(j)})$ are non-overlapping.  

The next important step is adoption of the weight function $g_{s\,l_{nA}\,J_{F}}^{(j)}(k_{nA})$.  In \cite{thompson} two different prescriptions were used for the weight function for resonant and non-resonant bins.  We use  for the non-resonant bins
\begin{align}
g_{s\,l_{nA}\,J_{F}}^{(j)}(k_{nA}) \,= \,e^{-i\,\delta_{s\,l_{nA}\,J_{F}}(k_{nA})}
\label{weightfunct1}
\end{align}
and for the resonance bin
\begin{align}
g_{s\,l_{nA}\,J_{F}}(k)=e^{i\delta_{s\,l_{nA}\,J_{F}}(k)}\sin(\delta_{s\,l_{nA}\,J_{F}}(k)),
\label{weightresbin1}
\end{align}
where $\delta_{s\,l_{nA}\,J_{F}}(k_{nA})$  is the $n-A$  scattering phase shift. 

The radial scattering wave function  $\,u_{k_{nA}s\,l_{nA}\,,J_{F}}^{(j)(+)}(r_{nA})\,$  should describe the resonance scattering  in the bin covering the resonant region.  
In the  $R$-matrix  approach the coordinate space over $r_{nA}$ is divided into the internal, $r_{nA} \leq R_{nA}$, and external, $r_{nA} > R_{nA}$, regions. 
In the internal region in the one level approximation
\begin{align}
&u_{k_{nA}\,s\,l_{nA}\,J_{F}}^{(int)}  = -i\,\sqrt{\frac{k_{nA}}{\mu_{nA}}} \,
e^{-i\,\delta_{l_{nA}}^{hs}}\, \frac{[\Gamma_{nA\,s\,l_{nA}\,J_{F}}(E_{nA})]^{1/2}}{E_{R} - E_{nA} - i\,\Gamma_{nA\,s\,l_{nA}\,J_{F}}(E_{nA})/2}\,X_{int}. 
\label{usinresint1}
\end{align}
Here,  $\Gamma_{nA\,s\,l_{nA}\,J_{F}}(E_{nA})$ is the  partial resonance width  in the channel $n + A$,  $\,\delta_{sl_{nA}}^{hs}$ is the hard sphere scattering phase shift,  $\,X_{int}\,$  is  an
eigenfunction of the  Hamiltonian describing the compound system $F=n+A$.
At the  channel radius $r_{nA}= R_{nA}$
\begin{align}
X_{int}= \frac{1}{R_{nA}}\,\sqrt{2\,\mu_{nA}\,R_{nA}}\,\gamma_{s\,l_{nA}\,J_{F}},
\label{Xint1}
\end{align}
where $\gamma_{s\,l_{nA}\,J_{F}}$  is the reduced width amplitude in the channel with quantum numbers $\,s,\,l_{nA}\,$ and $\,J^{F}$.
In the external region ($r_{nA} > R_{nA}\,$)  in the representation with a given channel spin  $s$ and  orbital angular  momentum $l_{nA}$ wave function $\,u_{k_{nA}s\,l_{nA}\,J_{F}}^{(j)(+)}(r_{nA})\,$  takes the standard form  
\begin{align}
&u_{k_{nA}\,s\,l_{nA}\,J_{F}}^{(ext)( + )} = [I_{l_{nA}}(k_{nA},\,r_{nA})-  S_{nA\,s\,l_{nA};\,nA\,s\,l_{nA}}^{J_{F}}\,O_{l_{nA}}(k_{nA},\,r_{nA})] ,                                              
\label{uelnA1}
\end{align}
where $I_{l_{nA}}(k_{nA},\,r_{nA})$ and $O_{l_{nA}}(k_{nA},\,r_{nA})$  are incoming and outgoing spherical waves, respectively.  By equating the internal $\,u_{k_{nA}\,s\,l_{nA}\,J_{F}}^{(int)}\,$ and external $\,u_{k_{nA}\,s\,l_{nA}\,J_{F}}^{(ext)( + )}$\,  wave functions at the channel radius $r_{nA}=R_{nA}$  we get an expression for the resonant $S$ matrix elastic scattering element $S_{nA\,s\,l_{nA};\,nA\,s\,l_{nA}}^{J_{F}}$, which  at energies near the resonances  takes the form
\begin{align}
&S_{nA\,s\,l_{nA};\,nA\,s\,l_{nA}}^{J_{F}}= e^{-2\,i\,\delta_{s\,l_{nA}}^{hs}}\,\big(1+ i\,  \frac{\Gamma_{nA\,s\,l_{nA}\,J_{F}}(E_{nA})}{E_{R} - E_{nA} - i\,\Gamma_{nA\,s\,l_{nA}\,J_{F}}(E_{nA})/2}\big),
\label{elscatSmatrix2}
\end{align}
where $E_{R}$ is the real part of the  resonance energy. 
 $\,\delta_{s\,l_{nA}}^{hs}\,$ is the hard-sphere scattering phase shift in the channel $n + A$  determined by  equation
\begin{align}
e^{-2\,i\,\delta_{l_{nA}}^{hs}}\,= \, \frac{I_{l_{nA}}(k_{nA},\,R_{nA})}{O_{l_{nA}}(k_{nA},\,R_{nA})}.
\label{hsscphshift1}
\end{align}
Thus in the external region $u_{k_{nA}\,s\,l_{nA}\,J_{F}}^{(ext)( + )}$ can be expressed in terms of the observable partial resonance widths and  resonance energies. 

Another possible approach is the potential one.  In the potential approach first we introduce the overlap function $I_{A\,{\rm {\bf k}}_{nA}}^{F(-)*}= <\psi_{F\,{\rm {\bf k}}_{nA}}^{(-)}|\varphi_{A}>$,  
where $\varphi_{A}$  is the bound state wave function of nucleus $A$ and  $\psi_{F}^{(-)}$ is the eigenfunction of the continuum spectrum of the Hamiltonian $\,H= K_{nA} + V_{nA} + H_{A}$ of the system $F=n+A$.  This overlap function is  approximated  as  \cite{bunakov1970}
\begin{align}
I_{A\,{\rm {\bf k}}_{nA}}^{F(-)*}= S_{A}^{F}\,u_{{\rm {\bf k}}_{nA}}^{(-)*},
\label{overlapsnglpartapprox1}
\end{align}
where $S_{A}^{F}$  is the spectroscopic factor  of the configuration $n+ A$  in $F$ and  $u_{{\rm {\bf k}}_{nA}}^{(-)*}$   is a solution of the Schr\"odinger equation
\begin{align}
(E_{nA} - K_{nA} - V_{nA}^{sp})\,u_{{\rm {\bf k}}_{nA}}^{(-)*}({\rm {\bf r}}_{nA})=0.
\label{SchequnA1}
\end{align}
The external part of the single-particle  wave function  $u_{k_{nA}\,s\,l_{nA}\,J_{F}}^{(ext)( + )}$ (we recovered here the spins)   is given by Eq. (\ref{uelnA1})  where the elastic scattering $S$ matrix  is generated by the potential $V_{nA}^{sp}$.  This $S$ matrix element in the single-particle model is given by
\begin{align}
&S_{nA\,s\,l_{nA};\,nA\,s\,l_{nA}}^{(sp)\,J_{F}}= e^{-2\,i\,\delta_{s\,l_{nA}\,J_{F}}^{sp}}\,\big(1+ i\,  \frac{\Gamma_{nA\,s\,l_{nA}\,J_{F}}^{sp}}{E_{R} - E_{nA} - i\,\Gamma_{nA\,s\,l_{nA}\,J_{F}}^{sp}/2}\big),
\label{elscatSmatrixsp1}
\end{align}
where $\delta_{s\,l_{nA}\,J_{F}}^{sp}$ is the potential non-resonance scattering phase shift,
 $\Gamma_{nA\,s\,l_{nA}\,J_{F}}^{sp}$ is the single-particle neutron  resonance width. Then the observable resonance width  is written as
\begin{align}
\Gamma_{nA\,s\,l_{nA}\,J_{F}}= S_{A}^{F}\,\Gamma_{nA\,s\,l_{nA}\,J_{F}}^{sp}.
\label{obsspreswidth1}
\end{align}
 
Now we return to the prior form of the stripping amplitude.  
After defining the CDCC wave function in the final state it is clear from Eqs (\ref{ShreqCDCCf1})  and  
\begin{align}
(E - K - V_{pn}  - U_{dA})\,\varphi_{pn}\,\chi_{dA}^{(+)}=0
\label{SchreqPhii1}
\end{align}
that  amplitudes   (\ref{CDCCvolconstr1})  and   (\ref{surfCDCCconstr1})  coincide.  The main advantage of the surface amplitude (\ref{surfCDCCconstr1})  is that the convergence problem 
for the stripping to resonance is solved because the integration over $\rho_{dA}$ is taken at  the finite  $\rho_{dA}= {\mathcal R}_{dA}$  and the integration over $r_{pn}$ is constrained owing to the presence of $\varphi_{pn}$.  Because of these two integrations  the contribution of the peripheral region over $r_{nA}$  in the surface matrix element is enhanced compared to the conventional volume matrix element. But the surface matrix element is not fully peripheral over $r_{nA}$ because of the large non-locality of the prior amplitude (typically 20-25 fm). It means that  small $\rho_{pF}$, and, correspondingly, small $r_{nA}$ can contribute making  non-peripheral contribution also possible, especially when the energy increases.  Eq. (\ref{surfCDCCconstr1}) is the main result of our paper.  

There is  one more point about CDCC to discuss.  We have assumed that the CDCC wave function given by Eq. (\ref{cdccwf1})  is a solution of Eq. (\ref{ShreqCDCCf1}).  As we have discussed the constraint imposed on $ l_{nA}^{max}$ allows us to diminish the role of the rearrangement channels. However, it may not be enough and a more sophisticated  truncation procedure is achieved by using the projector
\begin{align}
{\hat P}_{nA}= \sum\limits_{l_{nA}=0}^{l_{nA}^{max}}\,\sum\limits_{m_{l_{nA}}=-l_{nA}}^{l_{nA}}\,\int\,{\rm d}\Omega_{{\rm {\bf r}}_{nA}}\,
Y_{l_{nA}\,m_{l_{nA}}}({\rm {\bf{\hat r}}}_{nA})\,Y_{l_{nA}\,m_{l_{nA}}}^{*}({\rm {\bf {\hat r}}}_{nA}').
\label{PpnprojectornA1}
\end{align}

Applying the projector $P_{nA}$  to Eq. (\ref{ShreqCDCCf1})  from the right we get  the Schr\"odinger equation for the  CDCC wave function in the final state in the projected model space:
\begin{align}
\Psi_{(P_{nA})\,f}^{CDCC(-)*}\,(E- {\overleftarrow K} -  U_{pA}^{P_{nA}}- V_{pn}^{P_{nA}} - V_{nA}) =0, 
\label{projSchreqn1}
\end{align}
where   $\,\Psi_{(P_{nA})\,f}^{CDCC(-)*}= \Psi_{f}^{CDCC(-)*}\,P_{nA}$,    $\,U_{pA}^{P_{nA}}= P_{nA}\,U_{pA}\,P_{nA}\,$
and  $V_{pn}^{P_{nA}}=P_{nA}\,V_{pn}\,P_{nA}$.     Note that  the projector $P_{nA}$ acts on ${\rm {\bf r}}_{nA}$, hence, it modifies $U_{pA}$ and $V_{pn}$, which can be expressed in terms of the radii ${\rm {\bf r}}_{nA}$ and ${\bm \rho}_{pF}$.  The potential $\,V_{nA}\,$ remains intact to the action of $\,P_{nA}\,$ because it depends only on $r_{nA}$ rather than on ${\rm {\bf r}}_{nA}$. 

 In the projected model space the rearrangement channels are suppressed.  For example,  if  we add  to the CDCC wave function  the component of the rearrangemenet channel  $\,\varphi_{pn}\,\chi_{dA}^{(+)}\,$,  application to it of the projector $\,P_{nA}\,$    at $\,\rho_{dA} >> r_{pn}\,$   brings an additional suppression factor $\,\rho_{dA}^{-2}$ \cite{austern87}.  
In the projected model space the conventional volume matrix element and the matrix element in the surface integral formalism  do not coincide.  To show it  we go back   to the 
conventional volume matrix element  (\ref{CDCCvolconstr1}), in which we replace $\,\Psi_{f}^{CDCC(-)}\,$ by $\,\Psi_{(P_{nA})f}^{(-)}\,$ without changing potentials. Then we get                                                       
\begin{align}
&{\mathcal M}_{conv}^{CDCC(prior)} \,= <\Psi_{(P_{nA})\,f}^{CDCC(-)}\big| U_{pA} + V_{nA}^{sp}  -  U_{dA}\big| \varphi_{pn}\,\chi_{dA}^{(+)}>\Big|_{ \rho_{dA} \leq {\mathcal R}_{dA};\,   r_{pn} \leq {\mathcal R}_{pn} }.    
\label{CDCCvolproj1}                                                       
\end{align}
To transform this matrix element to the surface integral form we rewrite
\begin{align}
U_{pA} + V_{nA}^{sp}  -  U_{dA} = (U_{pA} - U_{pA}^{P_{nA}}) + [U_{pA}^{P_{nA}} +  V_{pn}^{P_{nA}} + V_{nA}^{sp}] - [V_{pn} + U_{dA}] + (V_{pn} - V_{pn}^{P_{nA}})    .
\label{transoper1}
\end{align}  
Taking into account the Schr\"odinger equations for $\,\Psi_{(P_{nA})f}^{CDCC(-)*}\,$ and $\,\varphi_{pn}\,\chi_{dA}^{(+)}\,$  we can replace  the bracketed operator $[U_{pA}^{P_{nA}} +  V_{pn}^{P_{nA}} + V_{nA}^{sp}] $   by $\,E- {\overleftarrow K}\,$   and  $[V_{pn} + U_{dA}]$ by $E - {\overrightarrow K}$.  
Then  Eq. (\ref{CDCCvolproj1})  can be reduced to
\begin{align}
&{\mathcal M}_{conv}^{CDCC(prior)} \,= <\Psi_{(P_{nA})\,f}^{CDCC(-)}\big| U_{pA} + V_{nA}^{sp}  -  U_{dA}\big| \varphi_{pn}\,\chi_{dA}^{(+)}>\Big|_{ \rho_{dA} \leq {\mathcal R}_{dA};\,   r_{pn} \leq {\mathcal R}_{pn} }  \\
&={\mathcal M}_{surf}^{CDCC(prior)}   +  M_{aux}^{CDCC(prior)}.
\label{CDCCsurfauxpr1}                                                       
\end{align}
Here the matrix element in the surface integral  representation is
\begin{align}
&{\mathcal M}_{surf}^{CDCC(prior)}  = <\Psi_{(P_{nA})\,f}^{CDCC(-)}\big|  {\overrightarrow K} -  {\overleftarrow K}  \big| \varphi_{pn}\,\chi_{dA}^{(+)}>\Big|_{ \rho_{dA} \leq {\mathcal R}_{dA};\,   r_{pn} \leq {\mathcal R}_{pn} }                                                                                        \nonumber\\
&= <\Psi_{(P_{nA})\,f}^{CDCC(-)}\big|  {\overrightarrow K}_{dA} -  {\overleftarrow K}_{dA}  \big| \varphi_{pn}\,\chi_{dA}^{(+)}>\Big|_{ \rho_{dA} \leq {\mathcal R}_{dA};\,   r_{pn} \leq {\mathcal R}_{pn} }                                                                                                                        \nonumber\\
&=  - \frac{R_{dA}^{2}}{2\,\mu_{dA}}\,\int\,{\rm d}{\rm {\bf  r}}_{pn}\,\varphi_{pn}({\rm {\bf r}}_{pn})\,\int\,{\rm d}\Omega_{ {\rm {\bm { \rho}}}_{dA}}                \Big[\Psi_{(P_{nA})f}^{CDCC(-)*}({\rm {\bm \rho}}_{dA},{\rm {\bf r}}_{pn})\frac{{\partial \chi_{dA}^{(+)}({\rm {\bm \rho}}_{dA})}}{{\partial {\rho_{dA}}}}      \nonumber\\                         
&- \chi_{dA}^{(+)}({\rm {\bm \rho}}_{dA})\frac{{\partial \Psi_{(P_{nA})\,f}^{CDCC(-)*}({\rm {\bm \rho}}_{dA},{\rm {\bf r}}_{pn})}}{{\partial {\rho_{dA
}}}}\Big]\Big|_{ \rho_{dA} ={\mathcal R}_{dA};\, r_{pn} \leq {\mathcal R}_{pn} },
\label{CDCCsurfpr1}
\end{align}
where we took into account that the matrix element from  ${\overleftarrow K}_{pn} - {\overrightarrow K}_{pn}$ vanishes. 
The auxiliary matrix element is given by
\begin{align}
&{\mathcal M}_{aux}^{CDCC(prior)}  = <\Psi_{(P_{nA})\,f}^{CDCC(-)}\big|  U_{pA} - U_{pA}^{P_{nA}} +  V_{pn} - V_{pn}^{P_{nA}}  \big| \varphi_{pn}\,\chi_{dA}^{(+)}>\Big|_{ \rho_{dA} \leq {\mathcal R}_{dA};\,   r_{pn} \leq {\mathcal R}_{pn} }                                                                       \nonumber\\                                                                                       
&= <\Psi_{(P_{nA})\,f}^{CDCC(-)}\big|  P_{nA}(U_{pA}  +  V_{pn} )Q_{nA}  \big| \varphi_{pn}\,\chi_{dA}^{(+)}>\Big|_{ \rho_{dA} \leq {\mathcal R}_{dA};\,   r_{pn} \leq {\mathcal R}_{pn} } .
\label{CDCCauxpr1}
\end{align}
To obtain  Eq. (\ref{CDCCauxpr1})   we took into account  that   $P_{nA} + Q_{nA}=1, \,\,P_{nA}^{2}=  P_{nA},\,\,$
$\,\Psi_{(P_{nA})\,f}^{CDCC(-)*}\,(U_{pA} - U_{pA}^{P_{nA}} +  V_{pn} - V_{pn}^{P_{nA}})\,= \,\Psi_{(P_{nA})\,f}^{CDCC(-)*}P_{nA}\,(U_{pA} - U_{pA}^{P_{nA}} +  V_{pn} - V_{pn}^{P_{nA}}  ) = \Psi_{(P_{nA})\,f}^{CDCC(-)}\,P_{nA}\,(U_{pA}  +  V_{pn}  )\,Q_{nA} $.  The potential $\,P_{nA}\,(U_{pA}  +  V_{pn}  )\,Q_{nA}\,$  couples  low  orbital angular momenta  $l_{nA}$  with the large $\,l_{nA}\,$ from the subspace $\,Q_{nA}$. Thus the auxiliary term adds a model dependence because by taking into account this term we  go beyond the limits of the model space constrained by the projector $P_{nA}$.

\section{Numerical Results}
\label{Calculations1}

In this section we present some calculations corroborating our theoretical findings although the code for the surface untegral formalism in the CDCC approach is not yet available and the work on it is in progress.

\subsection{Stripping to bound state.  Reaction  ${}^{14}{\rm C}(d,\,p){}^{15}{\rm C}(2s_{1/2}, E_{x}=0.0\,{\rm MeV}).$}

\begin{figure}
\includegraphics[scale=1.3]{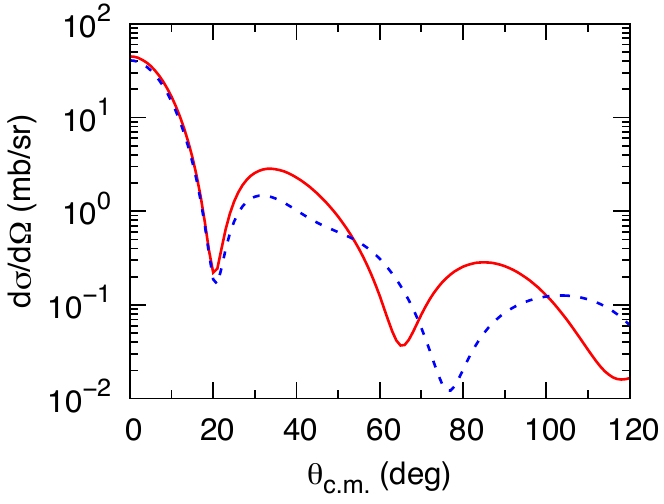}
\caption{(Color online) Prior DWBA  differential cross sections for the  ${}^{14}{\rm C}(d,\,p){}^{15}{\rm C}(gr. st.)$  at $E_{d}=23.4$ MeV.
Solid red line is obtained using the optical potential $U_{nA}$  when calculating $U_{dA}$;  blue dotted line is obtained with 
$U_{nA} = V_{nA}^{sp}$  in $U_{dA}$. } 
\label{fig_ImUnAAWBA1}
\end{figure}

\begin{figure}
\includegraphics[scale=1.3]{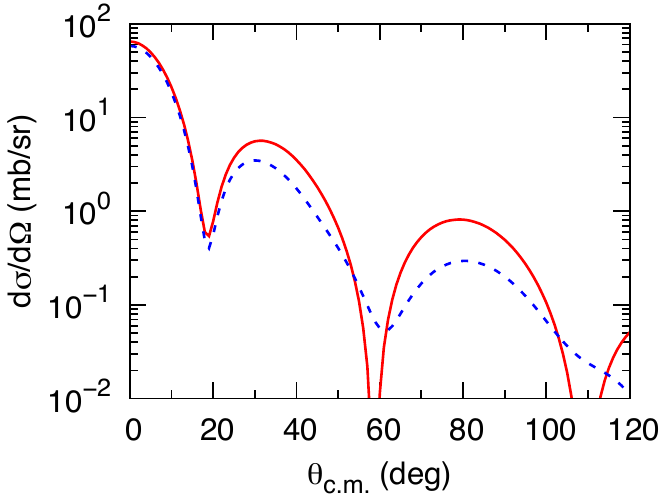}
\caption{(Color online) Prior CDCC  differential cross sections for the  ${}^{14}{\rm C}(d,\,p){}^{15}{\rm C}(gr. st.)$  at $E_{d}=23.4$ MeV.
Notations are the same as in Fig.  \ref{fig_ImUnAAWBA1}. } 
\label{fig_ImUnACDCC1}
\end{figure}

First we present the effect of the auxiliary matrix element  (\ref{CDCCaux2}) .  To this end we performed calculations using the prior DWBA  amplitude 
\begin{align}
{\mathcal  M}^{DW(prior)} = < \chi_{pF}^{(-)}\,\varphi_{nA}^{F}\big| U_{pA} + U_{nA} - U_{dA}\big|\varphi_{pn}\,\chi_{dA}^{(+)}>
\label{DWBAprbst1}
\end{align}
and the prior CDCC amplitude
\begin{align}
{\mathcal  M}^{CDCC(prior)} = < \Psi_{f}^{CDCC(-)}\big| U_{pA} + U_{nA} - U_{dA}\big|\varphi_{pn}\,\chi_{dA}^{(+)}>.
\label{CDCCprbst1}
\end{align}
In both amplitudes  to calculate the initial  distorted wave  $\chi_{dA}^{(+)}$ we use the optical potential $U_{dA}$ prescribed by the  ADWA  using  the zero-range Johnson-Sopper prescription  \cite{johnson70}) in which  the $d-A$ optical potential $U_{dA}$ is given by the sum $U_{pA}+ U_{nA}$ taken at $r_{pn}=0$ and at half of the deuteron incident energy .  
In all the calculations we use Koning-Delaroche potential \cite{koningdelroche2003} for the $N-A$ optical potentials. We use the spectroscopic factor $S_{A}^{F}=1$ for $n+{}^{14}{\rm C}$  configuration 
in the ground state of ${}^{15}{\rm C}$.
By comparing the differential cross sections obtained using the complex $U_{nA}$ and the real $U_{nA}= V_{nA}^{sp}$ 
we can estimate the effect of the auxiliary terms 
\begin{align}
{\mathcal M}_{aux}^{DW(prior)}=  < \chi_{pF}^{(-)}\,\varphi
_{nA}\big| {\rm Im}U_{nA}\big|\varphi_{pn}\,\chi_{dA}^{(+)}>
\label{prauxDWBA1}
\end{align}
and
\begin{align}
{\mathcal M}_{aux}^{CDCC(prior)}=  < \Psi_{f}^{CDCC(-)}\big| {\rm Im}U_{nA}\big|\varphi_{pn}\,\chi_{dA}^{(+)}>.
\label{prauxCDCC1}
\end{align}
Clearly our calculations give a rather qualitative estimation of the auxiliary term effect because  when we change the $U_{nA}$ in the transition operator we simultaneously change the distorted wave $\chi_{dA}^{(+)}$ in the initial state while in the auxiliary amplitudes (\ref{prauxDWBA1})  and
  (\ref{prauxCDCC1})  with changing $U_{nA}$  only the transition operator should change. Hence our calculations overestimate the effect of the auxiliary term. The calculations are done for the ${}^{14}{\rm C}(d,p){}^{15}{\rm C}(2s_{1/2}, E_{x}=0.0\,{\rm MeV})$ at the deuteron energy of $E_{d}= 23.4$ MeV.   The results are shown in  Figs.  \ref{fig_ImUnAAWBA1}  and  \ref{fig_ImUnACDCC1}.

The replacement in $U_{dA}$  of  the real potential $V_{nA}^{sp}$ by the complex optical potential $U_{nA}$ changes the differential cross section at forward angles by  $10\%$ for the DWBA and by $11\%$ for the CDCC.  

As we can see in Figs \ref{fig_ImUnAAWBA1} and  \ref{fig_ImUnACDCC1} the replacement of $U_{nA}$ by $V_{nA}^{sp}$ makes very little effect on the differential cross section in the region of the first stripping peak, confirming that at low energies the contribution from the nuclear interior is small at forward angles but increases with angle increasing. Hence, at low energies the replacement of $U_{nA}$ by $V_{nA}^{sp}$ does not affect
the spectroscopic information, like ANCs or spectroscopic factors, which is extracted from the normalization of the calculated differential cross section to the experimental one in the first stripping peak.  

\begin{figure}
\includegraphics[scale=1.3]{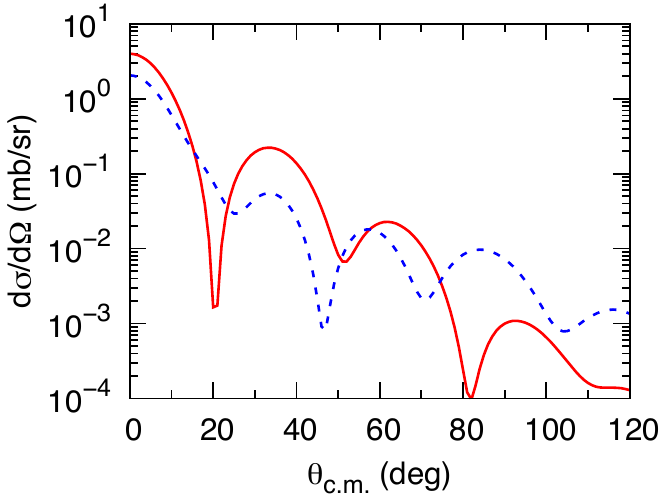}
\caption{(Color online) Prior DWBA  differential cross sections for the  ${}^{14}{\rm C}(d,\,p){}^{15}{\rm C}(2s_{1/2}, E_{x}=0.0\,{\rm MeV})$  at $E_{d}=60$ MeV.
Notations are the same as in Fig.  \ref{fig_ImUnAAWBA1}. } 
\label{fig_ImUnAAWBA60MeV1}
\end{figure}

Similar calculations  for $60$ MeV deuterons give quite different results.  In Fig.  \ref{fig_ImUnAAWBA60MeV1}
we present the prior DWBA differential cross sections for two different choices of the $n-A$ potential used to calculate $U_{dA}$. As we see the difference is quite significant but it comes mainly owing to the different initial distorted waves $\chi_{dA}^{(+)}$  generated by different $U_{dA}$. If for $E_{d}= 23.4$  MeV this difference was not important because the reaction was peripheral, it is not the case for $60$ MeV, when the deuteron stripping reaction is contributed also by the nuclear interior \cite{matt}. Unfortunately we are  not able to calculate the matrix element from ${\rm Im}U_{nA}$  without changing the initial distorted wave. 

\begin{figure}
\includegraphics[scale=1.3]{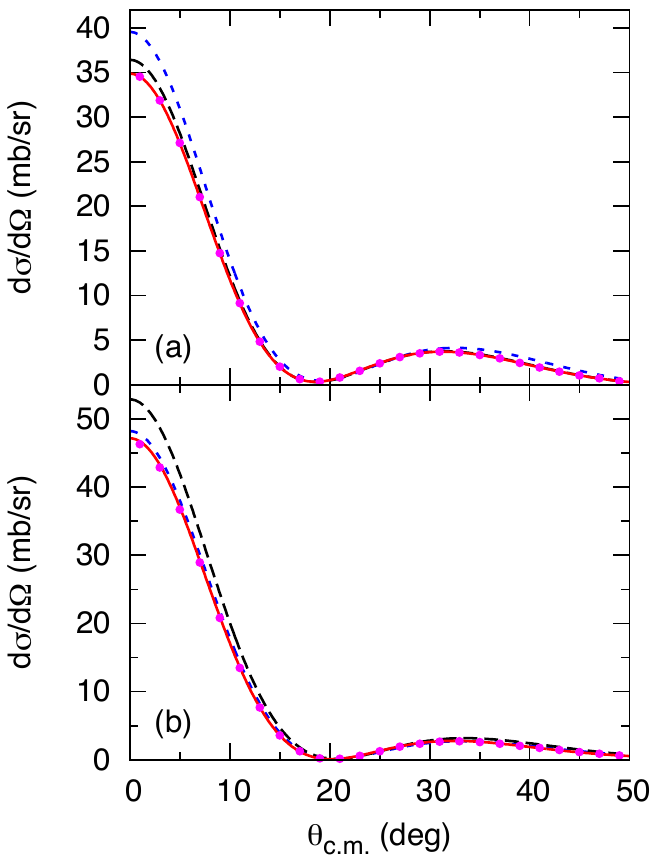}
\caption{(Color online)  Post (panel (a))  and prior (panel (b)) CDCC  differential cross sections for the  ${}^{14}{\rm C}(d,\,p){}^{15}{\rm C}(2s_{1/2}, E_{x}=0.0\,{\rm MeV})$  at $E_{d}=23.4$ MeV.  In the post  form (panel (a)) the cut-off is introduced  over the $p-n$  partial waves  in the continuum component of the initial  CDCC scattering wave function: $\,l_{pn}^{max}=0\,$ - red solid line, $l_{pn}^{max}= 2$  -  black dashed line, $l_{pn}^{max}=4$ - blue short dashed line,  $l_{pn}=6$ - dots ; in  the prior  form (panel (b)) the cut-off is introduced  over the $n-A$  partial waves  in the continuum component of the final  CDCC scattering wave function: $\, l_{nA}^{max}=1\,$ - red solid line,  $ \,l_{nA}^{max}= 2$  -  black dashed line,  $\,l_{nA}^{max}=3$ - blue short dashed line,  $l_{nA}=4$- dots.} 
\label{fig_postprior15C234MeV1}
\end{figure}

In the second type of  calculations we compared the post and prior CDCC amplitudes  for the  ${}^{14}{\rm C}(d,p){}^{15}{\rm C}(2s_{1/2}, E_{x}=0.0\,{\rm MeV})$ reaction at the deuteron energy of $E_{d}= 23.4$ MeV.   In Fig  \ref{fig_postprior15C234MeV1}  we compare the dependence  of the CDCC amplitudes  on the maximum $l_{pn}$ of the continuum $p-n$ states in the post form and maximum $l_{nA}$ of the continuum $n-A$ states in the prior form.  For both post and prior forms $l_{pn}=4$  and  $l_{nA}=4$, correspondingly, are enough to achieve convergence.

\begin{figure}
\includegraphics[scale=1.3]{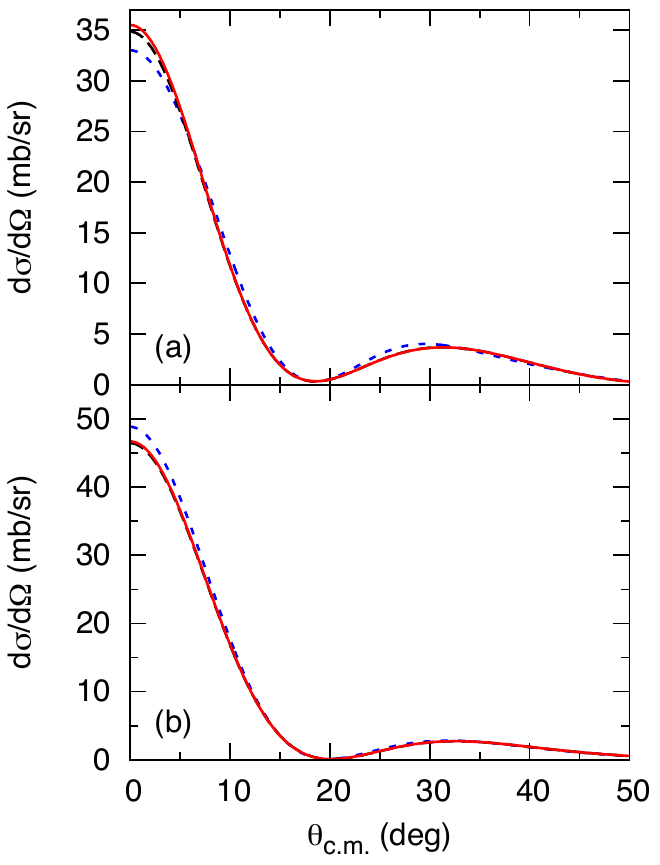}
\caption{(Color online) Convergence of the post (panel (a)) and prior (panel (b)) CDCC  differential cross sections for the  ${}^{14}{\rm C}(d,\,p){}^{15}{\rm C}(2s_{1/2}, E_{x}=0.0\,{\rm MeV})$  at $E_{d}=23.4$ MeV.
$R_{match}=20$ fm-red solid line;  $R_{match}= 30$ fm- black dashed line;  $R_{match}=40$ fm -  blue short dashed line. } 
\label{fig_postpriorCDCCRmatch15C234MeV1}
\end{figure}

Now in Fig \ref {fig_postpriorCDCCRmatch15C234MeV1}    we demonstrate the convergence of the post and prior  CDCC  differential cross sections for the  ${}^{14}{\rm C}(d,\,p){}^{15}{\rm C}(2s_{1/2}, E_{x}=0.0\,{\rm MeV})$  at $E_{d}=23.4$ MeV as functions of ${\mathcal  R}_{dA}$ and  ${\mathcal R}_{pF}$ . In the FRESCO code this corresponds to parameter $R_{match}$.  
The post form converges at $R_{match}=40$ fm while the prior form converges at $R_{match}=30$ fm,
 although the  post form has nonlocality range in the matrix element  $9$ fm versus $24$ fm in the prior  form.
These calculations  demonstrate that the integration volumes  over ${\bm \rho}_{dA}$  and  ${\bm \rho}_{pF}$ in the CDCC matrix elements are constrained.  

\begin{figure}
\includegraphics[scale=1.3]{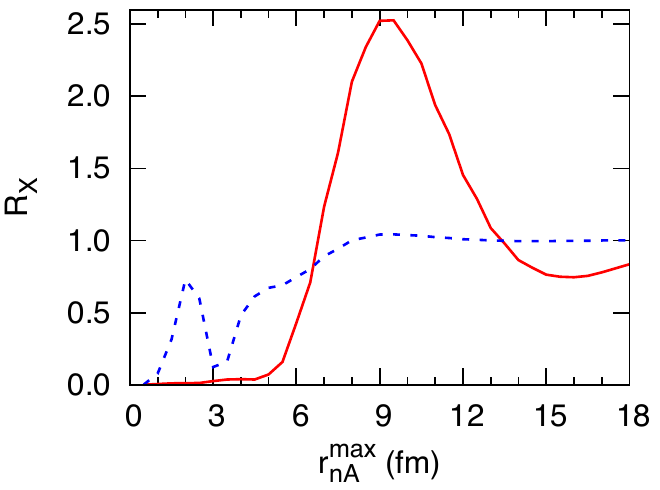}
\caption{(Color online) Dependence of the normalized post  CDCC  differential cross sections $\,R_{x}\,$ on $r_{nA}^{max}$ for the  $\,{}^{14}{\rm C}(d,\,p){}^{15}{\rm C}(2s_{1/2}, E_{x}=0.0\,{\rm MeV})\,$  at $E_{d}=23.4$ MeV.    $R_{x}$  is the ratio  of the  peak  CDCC differential cross section, in which the integral  over $r_{nA}$  is calculated up to  $r_{nA}^{max}$,  to the full  peak  CDCC differential cross section calculated at $r_{nA}^{max} \to \infty$.
Solid red line is the  normalized post CDCC form,  blue dotted line is the  normalized  prior CDCC fiorm.} 
\label{fig_postpriorCDCCrnA15C234MeV1}
\end{figure}

In Fig. \ref{fig_postpriorCDCCrnA15C234MeV1}   we show the convergence of the post and prior CDCC differential cross sections  as  functions of  $r_{nA}$  for the  ${}^{14}{\rm C}(d,\,p){}^{15}{\rm C}(2s_{1/2}, E_{x}=0.0\,{\rm MeV})$  at $E_{d}=23.4$ MeV.   To this end we calculated  the post and prior  CDCC differential cross sections in which the integration over  $r_{nA}$   was cut at the upper limit $r_{nA}^{max}$. By increasing $r_{nA}^{max}$  we  can determine the convergence of the CDCC differential cross sections  as functions of $r_{nA}^{max}$.
The convergence over $r_{nA}$  is important because depending on ${\rm {\bf r}}_{nA}$  the overlap function $I_{A}^{F}$  is the only source of the spectroscopic information, which can be extracted from the deuteron stripping  reactions. In the case under consideration, owing to the small neutron binding energy  
$\,\varepsilon_{n\,{}^{14}{\rm C}}^{{}^{15}{\rm C}} =1.218$ MeV in ${}^{15}{\rm C}$,  we expect a very slow convergence of the  CDCC matrix elements.  Nevertheless,  our calculations demonstrate  that the  prior form  converges at  $r_{nA}  \approx 9$ fm, while the convergence of the post form is achieved at $r_{nA}  > 20$ fm.
 This advantage of the prior form may be not decisive for the stripping to  bound states but could be important for stripping to resonance states. 

\subsection{Stripping to resonance state. Reaction ${}^{16}{\rm O}(d,\,p){}^{17}{\rm O}(1d_{3/2})$.}
Now we  proceed to the  calculation of the stripping to a  resonance state. We select  the reaction
${}^{16}{\rm O}(d,\,p){}^{17}{\rm O}(1d_{3/2})$  at $E_{d}=36$ MeV populating a resonance state of energy
$E_{x}=5.085$ MeV, which corresponds to the resonance level at $0.94$ MeV.   In all the calcullations shown below we use the single-particle approach for the $n-A$ resonant scattering wave function
calculated  in the Woods-Saxon potential with the radial parameter $r_{0}=1.25$ fm and diffusseness
$a=0.65$ fm.

In the first calculation we  compare the post and prior  calculations  following the procedure developed in \cite{muk2011}.  The post and prior  ADWA and prior  CCBA  (coupled channel  Born approximation) are used for comparison.  The prior ADWA is the standard  prior DWBA  in which the initial
deuteron potential  is given by the sum of the  optical  $U_{PA}$ and $U_{nA}$ potentials calculated at half of the deuteron energy using the zero-range Johnson-Sopper prescription \cite{johnson70}.  In the CCBA the final state wave function can be derived from 
Eq. (\ref{cdccwf1}) . To do it  we use the partial wave expansion of the binned $n-A$ continuum scattering wave function leaving only the resonance partial wave $l_{nA}=2$. The adopted bin covers the resonance  region and $\chi_{ {\rm {\bf q}}_{pF}({\bf k}_{nA})  }^{(res)(-)}({\rm {\bm \rho}}_{pF})$ 
corresponding to the resonance bin  has asymptotically both incident and outgoing waves.   The continuum 
resonance wave function component is coupled with two bound states in ${}^{17}{\rm O}$:  the ground state $1d_{5/2}$ and the first excited state $2s_{1/2}$. These terms are given by the sum over $i=0,1$ in Eq.  (\ref{cdccwf1}).  Thus schematically we can write the final state wave function in CCBA as
\begin{widetext}
\begin{align}
&\Psi_{f}^{CDCC(-)}({\rm {\bm \rho}}_{pF},\,{\rm {\bf r}}_{nA}) = \,\varphi_{nA}^{(0)}({\rm {\bf r}}_{nA})\,\chi_{{\rm {\bf q}}_{pF}}^{(0)(-)}({\rm {\bm \rho}}_{pF})\, +  \varphi_{nA}^{(1)}({\rm {\bf r}}_{nA})\,\chi_{{\rm {\bf q}}_{pF}}^{(1)(-)}({\rm {\bm \rho}}_{pF})                              \nonumber\\
& + {\overline \psi}_{ {\rm {\bf  k}}_{nA},\,l_{nA}=3 }^{(res)(-)}({\rm {\bf r}}_{nA})\,\chi_{ {\rm {\bf q}}_{pF}({\bf k}_{nA})  }^{(res)(-)}({\rm {\bm \rho}}_{pF}).
\label{cdccwf12}
\end{align}
\end{widetext}
Here, for simplicity, we omitted spins. The radial and momentum spherical harmonics are absorbed into 
${\overline \psi}_{ {\rm {\bf  k}}_{nA} }^{(res)(-)}({\rm {\bf r}}_{nA})$.  The distorted waves 
$\chi_{{\rm {\bf q}}_{pF}}^{(0)(-)}({\rm {\bm \rho}}_{pF})$ and  $\chi_{{\rm {\bf q}}_{pF}}^{(1)(-)}({\rm {\bm \rho}}_{pF})$  have only outgoing waves.  

\begin{figure}
\includegraphics[scale=1.3]{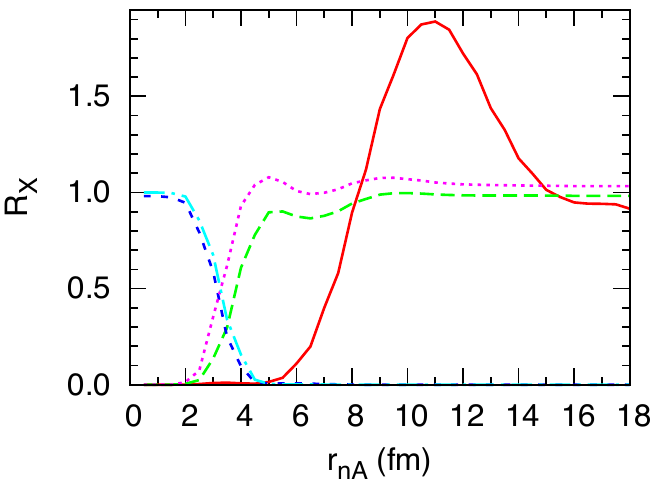}
\caption{(Color online) Dependence of the normalized  ADWA and CCBA  differential cross sections $\,R_{x}\,$  on  $\,r_{nA}\,$   for the deuteron stripping to resonance  $\,{}^{16}{\rm O}(d,\,p){}^{17}{\rm O}(1d_{3/2})\, $  at $\,E_{d}=36\,$ MeV.   Blue short and long dashed-dotted lines - the ratios $\,R_{X}\,$ of the  peak prior ADWA and CCBA differential cross sections, correspondingly, in which the  radial integral over $r_{nA}$  is  calculated  for  $r_{nA} \geq  r_{nA}^{min}$,  to the full differential cross section. 
Similarly, magenta dotted and green dashed lines are the ratios $\,R_{X}\,$ of the  peak prior ADWA and CCBA differential cross sections, correspondingly, in which the  radial integral over $r_{nA}$  is  calculated  in the inteval  $0 \geq  r_{nA} \leq  r_{nA}^{max}$,  to the full differential cross section.
 The red solid  line is the  $R_{X}$ dependence on $r_{nA}^{max}$ calculated for the post ADWA form.  Hence $r_{nA}$  on the abscissa  is   $\,r_{nA}^{min}$ for the blue short and long dashed lines  and $r_{nA}^{max}$ for the dotted magenta, dashed green  and solid red  lines .} 
\label{fig_Rx17O36MeVadwaccba}
\end{figure}

The results  of the calculations are shown in Fig.  \ref{fig_Rx17O36MeVadwaccba}.  Dependence of the peak  value of the normalized differential cross section $R_{X}$ on the $r_{nA}^{min}$ (blue short (ADWA) and long dashed-dotted (CCBA)  lines)  shows that 
the prior form  converges pretty fast being dominantly contributed by the region $r_{nA}  \lesssim 5$ fm
with following up small oscillations at larger $r_{nA}^{min}$. These small oscillations are better exposed 
on the ADWA  and  CCBA  lines, which show the dependence of the corresponding normalized  cross section on $r_{nA}^{max}$.
These oscillations practically disappear for $r_{nA}^{max} > 10$ fm, that is the prior form converges at $r_{nA}^{max}= {\mathcal R}_{nA}= 10$ fm.  Because in both ADWA and CCBA calculations the ADWA prescriptions was used,
the difference between both methods determines the effect of the coupling of the continuum resonant wave function in the final state with two bound states. As we see this effect is not significant.

\begin{figure}
\includegraphics[scale=1.3]{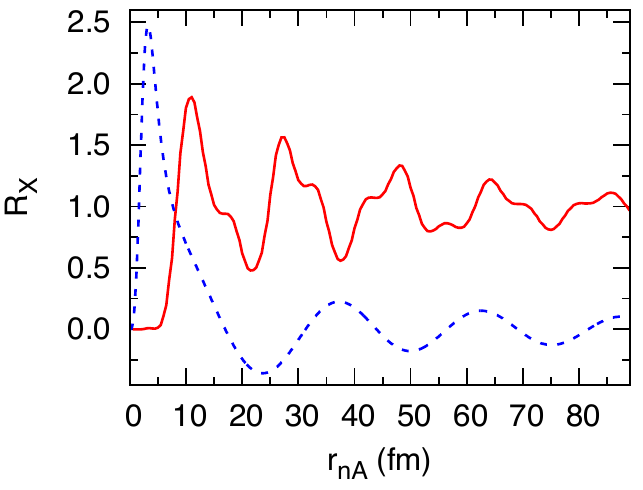}
\caption{(Color online) Solid red line- dependence on $r_{nA}^{max}$ of the normalized post  ADWA  differential cross section $\,R_{x}\,$ for the  stripping to resonance $\,{}^{16}{\rm O}(d,\,p){}^{17}{\rm O}(1d_{3/2})\,$  at $E_{d}=36$ MeV.
$R_{X}$ is calculated as the ratio of the ADWA differential cross section, in which the radial integral over $r_{nA}$  is  calculated  for  $0 \leq  r_{nA} \leq  r_{nA}^{max}$,  to the full differential cross section.  At each $r_{nA}^{max}$  the peak value of the differential cross section is used.  Blue dotted line- dependence on $r_{nA}$  of the binned radial resonant scattering wave function ${\overline \psi}_{k_{nA}\,s=1/2\,l_{nA}=2\, J_{F}=3/2}^{(res)}(r_{nA})$ .} 
\label{fig_ADWApostresconverg1}
\end{figure}

Meantime the post  form (solid red line) does not converge at much larger $r_{nA}^{max}$ sustaining significant oscillations even at  $r_{nA}^{max}  > 20$ fm.
To demonstrate a poor convergence of the post form  in Fig. \ref{fig_ADWApostresconverg1} we 
show the oscillation of the post ADWA  normalized  differential cross section $R_{X}$  as function 
of $r_{nA}^{max}$ (red solid line).  For comparison we show also the oscillation of the binned (the bin size is $1$ MeV)  resonant scattering waver function . As we  see, the oscillation of $R_{X}$ is caused  by the oscillation of the resonant scattering wave function.
Hence, the prior form  has evident advantage over the post one when dealing with the stripping to resonance. 

\begin{figure}
\includegraphics[scale=1.3]{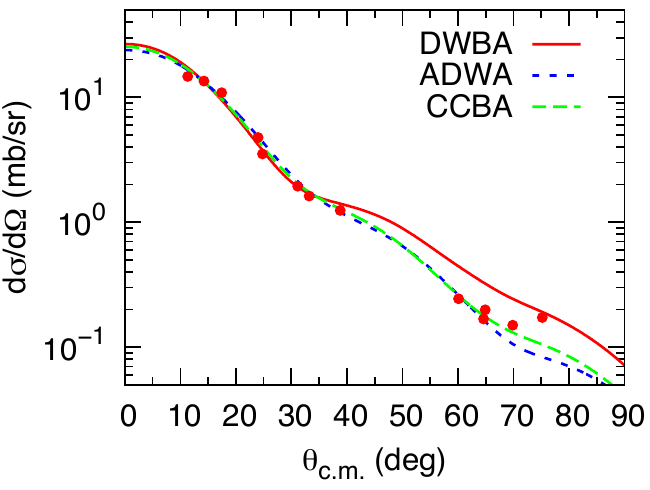}
\caption{(Color online) Angular distributions   for the deuteron stripping to resonance  $\,{}^{16}{\rm O}(d,\,p){}^{17}{\rm O}(1d_{3/2})\, $  at $\,E_{d}=36\,$ MeV.   Red solid line is the DWBA,  blue short dashed line is the ADWA,  green dashed line is the CCBA . All the angular distributions are normalized in the  region of the forward peak to the experimental one  -red dots \cite{cooper74}  .} 
\label{fig_angdistrres1}
\end{figure}

In Fig  \ref{fig_angdistrres1} the  angular distributions for the reaction $\,{}^{16}{\rm O}(d,\,p){}^{17}{\rm O}(1d_{3/2})\,$  at $E_{d}=36$  using prior DWBA, ADWA and CCBA are shown.  The CCBA, as explained above,  takes into account the coupling of the final resonant scattering wave function with the ground and first excited states 
in ${}^{17}{\rm O}$. As we can see the effect of coupling with the bound states has little effect on the angular distributions.
In the single-particle potential approach for the resonant scattering wave function the normalization of the theoretical cross section to the experimental one determines the spectroscopic factor, see Eq. (\ref{overlapsnglpartapprox1}).
From the normalization of the calculated differential cross sections we determined the spectrocopic factors: 
$SF= 0.89$ for the DWBA,   $SF=0.66$ for the ADWA  and $SF=0.73$ for the CCBA.  Using the single-particle 
neutron partial resonance width $\Gamma_{sp}=128$ keV, we get for the observable neutron widths 
$\Gamma_{n}=113.9$ keV for the DWBA,  $\Gamma_{n}= 84.5$ keV for the ADWA and $\Gamma_{n}=93.4$ keV for the CCBA.  The observed experimental value is $\Gamma_{n}=96 \pm 5$ keV. Thus the prior CCBA and  ADWA can be used to determine the observable partial resonance widths.

\begin{figure}
\includegraphics[scale=1.3]{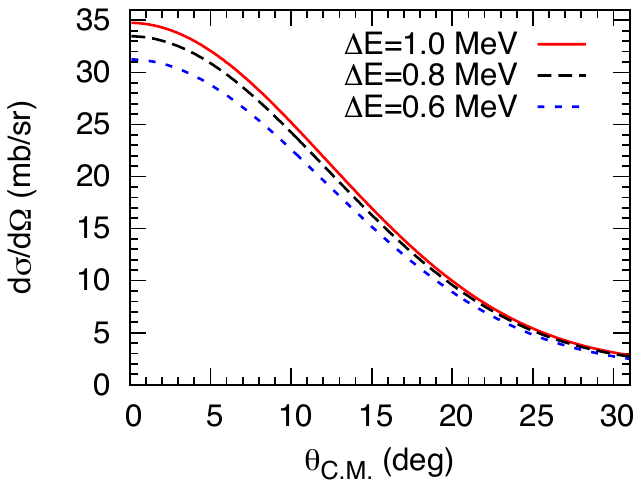}
\caption{(Color online) Angular distributions  for the deuteron stripping to resonance  $\,{}^{16}{\rm O}(d,\,p){}^{17}{\rm O}(1d_{3/2})\, $  at $\,E_{d}=36\,$ MeV calculated using prior CCBA for three different bins:  $1$ MeV-  red solid line, $0.8$ MeV- dashed  black line, $0.6$ MeV-  short dashed blue line.  } 
\label{fig_binsizedependangdistr1}
\end{figure}

Until now we have not  discussed the impact of the resonant bin width. In all the calculations shown above we used the bin width of $1$ MeV.  To check the impact of the bin width we performed  prior CCBA calculations with  three different bin widths. The results are shown in Fig \ref{fig_binsizedependangdistr1}.
The difference in the normalization of the  CCBA calculated differential cross sections at $1$ and $0.8$ MeV is only $3.7\%$. 

\begin{figure}
\includegraphics[scale=1.3]{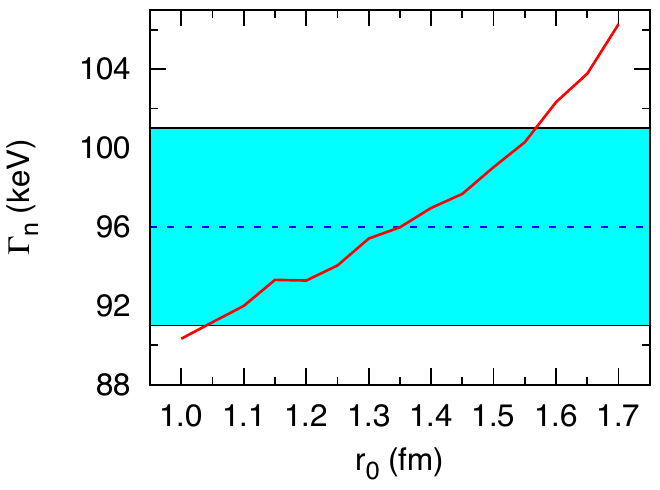}
\caption{(Color online) Solid red line- dependence on $r_{0}$ of the neutron resonance width extracted from the CCBA calculations of the $\,{}^{16}{\rm O}(d,\,p){}^{17}{\rm O}(1d_{3/2})\,$ reaction at $E_{d}=36$ MeV.
The blue dashed line is the experimental neutron resonance width of the $1d_{3/2}$ resonance in ${}^{17}{\rm O}$
and the blue strip is the resonance width's experimental uncertinty.} 
\label{fig_17Ogammar0dep}
\end{figure}

In our final calculations presented in Fig \ref{fig_17Ogammar0dep}we check the dependence of the extracted neutron resonance width on the radius $r_{0}$ of the $n-A$ Woods-Saxon potential, which supports the resonance state  $1d_{3/2}$.  This test is important for corroboration of our theoretical findings  and  shows how peripheral the deuteron stripping to resonance is. At each $1.0 \leq r_{0} \leq 1.7$  we calculated the CCBA differential cross section, normalized it to the exterimental one in the stripping peak in the angular distribution  and  determined  the spectroscopic factor, which is the normalization factor. 
For each $r_{0}$  from the derivative of the calculated scattering phase shift we determine the single-particle neutron resonance width and multiplying it by the  determined spectroscopic factor we find the observable resonance width 
shown in Fig. \ref{fig_17Ogammar0dep}.  
As we can see the determined neutron resonance width $\Gamma_{n}$  varies with variation of $r_{0}$  in the realistic interval $1.0-1.6$ fm  by $\pm 7\%$  from the experimental value of $96$ keV.   

\begin{figure}
\includegraphics[scale=1.3]{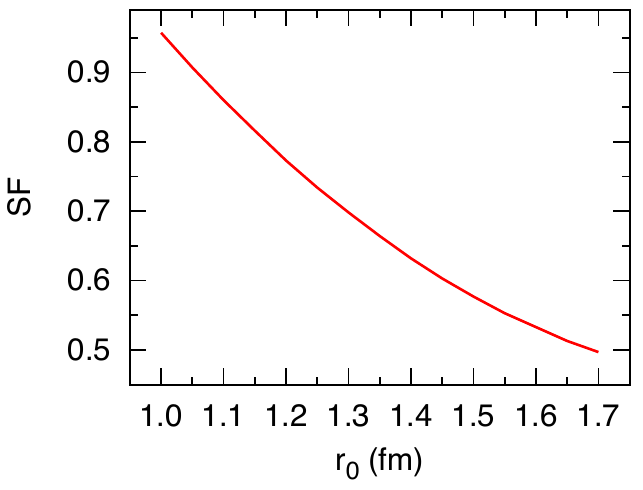}
\caption{(Color online) Solid red line- dependence on $r_{0}$ of the spectroscopic factor extracted from the CCBA calculations of the $\,{}^{16}{\rm O}(d,\,p){}^{17}{\rm O}(1d_{3/2})\,$ reaction at $E_{d}=36$ MeV.  } 
\label{fig_SFr01}
\end{figure}

The reaction is not peripheral  and this is clearly demonstrated by the $r_{0}$ dependence of $\Gamma_{n}$. In the case of the completely peripheral reaction  the  extracted $\Gamma_{n}$ should show none or a very little dependence on $r_{0}$.
From Fig. \ref{fig_17Ogammar0dep}  we can determine the radial parameter $r_{0}=1.35$  fm at which the extracted width coincides with the exeprimental one.  In Fig. \ref{fig_SFr01} we show the $r_{0}$ dependence of the spectroscopic 
factor.  Clearly the dependence on $r_{0}$  of the spectroscopic factor is much stronger than for $\Gamma_{n}$.
Taking into account that at $r_{0}=1.35$ fm the calculated $\Gamma_{n}$ coincides with the experimental one 
we can deternmine the spectroscopic factor  to be $S_{A}^{F}= 0.66 ^{+0.25}_{- 0. 1}$.

\section{Summary}
The goal of this paper was to develop a theory of the deuteron stripping to resonances  based  on the  surface-integral  formalism. First we demonstrated how the surface integral formalism worked for the deuteron stripping to bound states in the three-body model and then we considered a more realistic problem  in which a composite structure  of target nuclei is taken into account via optical potentials.  We explored different choices of channel wave functions and transition operators and showed that  the conventional CDCC volume matrix element  can be written in terms of the surface-integral matrix element, which is peripheral, and  the auxiliary matrix element, which determines the contribution of the nuclear interior over the variable $r_{nA}$.  This auxiliary matrix element appears  owing to the inconsistency in treating of the $n-A$ potential: this potential should be real in the final state to support  bound states or resonance scattering and  complex in the initial state to describe $n-A$ scattering.  

Our main result is a formulation of the  theory of the stripping to resonance states using the prior form of the surface integral formalism and CDCC method.  It is demonstrated that  the conventional CDCC volume matrix element coincides with the surface matrix element, which converges for the stripping to the resonance state. Also the surface representation (over the variable ${\rm {\bf r}}_{nA}$) of the stripping matrix element enhances the peripheral part of the amplitude although the internal contribution doesn't disappear  and increases with increase of the deuteron energy. 

Although the code for  the surface-integral formalism in the CDCCC approach is not yet available, we presented many calculations corroborating our findings both for the stripping to the bound state and the resonance. 
For the stripping to the bound state we use  ${}^{14}{\rm C}(d,\,p){}^{15}{\rm C}$ at  23.4 and 60 MeV of the deuteron incident energy.  It is shown how the contribution of the auxiliary term changes with energy. 
For the stripping to resonance state we explore ${}^{16}{\rm O}(d,\,p){}^{17}{\rm O}(1d_{3/2})$ reaction at  $E_{d}=36$ MeV.  Because the CDCC code for stripping to resonance is not yet available we use the CCBA and demonstrate 
that the prior form converges while the post form oscillates even at large distances.  We demonstrate how the resonance width can be extracted  from the analysis of the deuteron stripping to the resonance state.

\appendix

\section{Post-prior  DWBA discrepancy due to the $n-A$ potential inconsistency}
\label{VnAUnADWBA1}  

Here we  show  how the  inconsistency in the treatment of the $n-A$  potential  leads to the post-prior discrepancy of the DWBA 
amplitude. To this end we start from the post DWBA amplitude 
\begin{align}
{\mathcal M}^{DW(post)} = <\chi_{pF}^{(-)*}\,\varphi_{nA}|\Delta\,V_{pF}|\varphi_{pn}\,\chi_{dA}^{(+)}>
\label{DWpost1}
\end{align}
and derive from it the prior DWBA form. Here  
\begin{align}
\Delta V_{pF} = U_{pA} + V_{pn} - U_{pF}
\label{DeltaVpF1}
\end{align}
is the potential transition operator in the post form.  Let us take into account  Schr\"odinger equations  for the initial and  final channel wave functions
\begin{align}
(E- K - V_{pn} -  U_{dA})\,\varphi_{pn}\,\chi_{dA}^{(+)}=0
\label{SchreqPhii1}
\end{align}
and
\begin{align}
(E-K- V_{nA}^{sp} - U_{pF})\,\varphi_{nA}\,\chi_{pF}^{(-)*}=0.
\label{SchreqPhif1}
\end{align}

Then  Eq. (\ref{DWpost1})   can be transformed  into
\begin{align}
&{\mathcal M}^{DW(post)} = <\chi_{pF}^{(-)*}\,I_{A}^{F}|\Delta\,V_{pF}|\varphi_{pn}\,\chi_{dA}^{(+)}>    \nonumber\\
&= <\chi_{pF}^{(-)*}\,I_{A}^{F}| U_{pA}+ V_{nA}^{sp}  - U_{dA} - [V_{nA}^{sp}  + U_{pF}]+ [ V_{pn} + U_{dA}]|\varphi_{pn}\,\chi_{dA}^{(+)}>          \nonumber\\
&=  <\chi_{pF}^{(-)*}\,I_{A}^{F}| U_{pA}+ V_{nA}^{sp}  - U_{dA} + [E - {\overrightarrow K }  -V_{nA}^{sp}  - U_{pF}]|\varphi_{pn}\,\chi_{dA}^{(+)}>                                            \nonumber\\ 
&=  <\chi_{pF}^{(-)*}\,I_{A}^{F}| U_{pA}+ V_{nA}^{sp}  - U_{dA} + [E - {\overleftarrow K }  -V_{nA}^{sp}  - U_{pF}]|\varphi_{pn}\,\chi_{dA}^{(+)}>                                                                                  \nonumber\\
&= \, <\chi_{pF}^{(-)*}\,I_{A}^{F}| U_{pA}+ V_{nA}^{sp}  - U_{dA}|\varphi_{pn}\,\chi_{dA}^{(+)}>               \nonumber\\
&= {\mathcal  M}^{DW(prior)}.
\label{postpriorDWBA1}
\end{align}
Here we took into account  that the bracketed operators are the potentials in Eqs (\ref{SchreqPhii1})  and (\ref{SchreqPhif1}). 
Also because the matrix element contains   $\varphi_{pn}$ and $\varphi_{nA}$  the kinetic energy operator ${\overrightarrow K}$ can be transformed into ${\overleftarrow K}$.  Thus if we use the  real $V_{nA}^{sp}$  potential, which  generates the final bound state $(n\,A)$,  as the $n-A$ potential in the transition operator in the prior DWBA amplitude,  the post and prior DWBA  amplitudes coincide.  We note that in the proof of the equality of the post and prior forms we used the same $V_{nA}^{sp}$ potential both in the Schr\"odinger equation for the final state bound state wave function and in the transition operator of the prior form.  However,  the often used  global  optical potential  $U_{dA}$  is contributed by both $U_{pA}$ and $U_{nA}$ optical potentials. Similarly in the AWBA $U_{dA}$ is given by the sum of  $U_{pA} + U_{nA}$   with the $N-A$ optical potentials taken at half deuteron energy.
If we adopt $U_{nA}$  in the prior form transition operator  rather than $V_{nA}^{sp}$, then  the post and prior form DWBA amplitudes differ by the auxiliary amplitude\
\begin{align}
{\mathcal M}^{DW(post)} = {\mathcal  M'}^{DW(prior)}   +   {\mathcal M}_{aux}^{DW},
\label{postpriorauxDW1}
\end{align}	
where the prior DWBA amplitude is given now by
\begin{align}
{\mathcal M'}^{DW(prior)} = <\chi_{pF}^{(-)}\,I_{A}^{F}|U_{pA} +  U_{nA} - U_{dA}|\varphi_{pn}\,\chi_{dA}^{(+)}>   
\label{DWpriormod1}
\end{align}
and 
\begin{align}
{\mathcal M}_{aux}^{DW(prior)}  = <\chi_{pF}^{(-)}\,I_{A}^{F}\big|  U_{nA} - V_{nA}^{sp}  \big| \varphi_{pn}\,\chi_{dA}^{(+)}>.
\label{auxDW1}
\end{align}
In a modified  prior DWBA amplitude the transition operator  contains  the optical potential $U_{nA}$  rather than the real potential 
$V_{nA}^{sp}$  in the conventional prior form (\ref{postpriorDWBA1}). 
Thus the post and  prior DWBA amplitudes differ if we replace $V_{nA}^{sp}$  by  $U_{nA}$  in the  transition operator of the prior form, meaning that the inconsistency in the treatment of the $n-A$ potential leads to the post-prior discrepancy. If  we adopt 
${\rm Re}\,U_{nA}= V_{nA}^{sp}$   then
\begin{align}
{\mathcal M}_{aux}^{DW(prior)}  = <\chi_{pF}^{(-)}\,I_{A}^{F}\big|  {\rm Im}\,U_{nA} \big| \varphi_{pn}\,\chi_{dA}^{(+)}>.
\label{auxDW2}
\end{align}

\acknowledgments
C. A. B. and A.M. M. acknowledge that this material is based upon their work  supported by the U.S. Department of Energy, Office of Science, Office of  Nuclear Science.  C. B. A. is supported by the Award Numbers  DE-SC004971, DE-FG02-08ER41533.
A. M. M. is supported by the Award Numbers DE-FG02-93ER40773, DE-SC0004958. A.M.M. also acknowledges the support by the
DOE-NNSA under the Award Number  DE-FG52-09NA29467.   C. A. B. and A. M. M.  also acknowledge the support by  the US National Science Foundation under Award PHY- 1415656. A. S. K. acknowledges the support by the Australian Research Council.  DYP appreciates the warm hospitality he received during his visits to the Cyclotron Institute, Texas A\&M University. DYP gratefully acknowledges the supports of the National Natural Science Foundation of China (Grant Nos. 11275018 and 11035001) and the Chinese Scholarship Council (Grant No. 201303070253).

\end{document}